%% file: main.tex
\documentclass[a4paper, amsfonts, amssymb, amsmath, reprint, showkeys, nofootinbib, oneside,pdftex,colorlinks=true, linkcolor=blue, citecolor=blue, urlcolor=black]{revtex4-1}
\usepackage[english]{babel}
\usepackage[utf8]{inputenc}
\usepackage[colorinlistoftodos, color=green!40, prependcaption]{todonotes}
\usepackage{amsmath}
\usepackage{slashed}
\input{preamble}
\usepackage[pdftex, pdftitle={Article}, pdfauthor={Author}]{hyperref} 
\usepackage{geometry}
\begin{document}
\title{Aspects of Entanglement Entropy in $3d$ $\mathcal{N}=2$ SCFTs}

\author{Pedro Vicente Marto}
    \email{p.vicentemarto@uu.nl}
    \affiliation{Institute for Theoretical Physics, Utrecht University, 3584 CE Utrecht, The Netherlands}
    
\author{Umut Gürsoy}
    \email{u.gursoy@uu.nl}
    \affiliation{Institute for Theoretical Physics, Utrecht University, 3584 CE Utrecht, The Netherlands}

\author{Guim Planella Planas}
    \email{g.planellaiplanas1@uu.nl}
    \affiliation{Institute for Theoretical Physics, Utrecht University, 3584 CE Utrecht, The Netherlands}

\begin{abstract}
We investigate entanglement entropy in $3d$ $\mathcal{N}=2$ superconformal field theories from two different perspectives. We first confirm that the dependence of supersymmetric entanglement entropy (as defined in \cite{SUSYRenyi_Nishioka_2013}) on the entangling region is purely topological. We then compute entanglement entropy of $3d$ $\mathcal{N}=2$ superconformal gauge theories at arbitrary Yang-Mills coupling using the heat kernel method. This reveals a particular divergence originating in charge fluctuations in the vacuum, hence carrying information about internal symmetries of the theory. This computation leads to universal contributions to entanglement entropy from chiral and vector multiplets for unitary gauge groups of arbitrary rank, and has the advantage that the integral over the gauge moduli is simple to compute.
\end{abstract}


\maketitle

\input{sections/section01-UG}  

\appendix
\input{sections/appendix1}
\input{sections/appendix2}
\input{sections/appendix3}
\input{sections/appendix4}
\bibliography{references}
\end{document}

%% file: preamble.tex
\usepackage{amsthm}
\usepackage{mathtools}
\usepackage{physics}
\usepackage{xcolor}
\usepackage{graphicx}
\usepackage[left=23mm,right=13mm,top=35mm,columnsep=15pt]{geometry} 
\usepackage{adjustbox}
\usepackage{placeins}
\usepackage[T1]{fontenc}
\usepackage{lipsum}
\usepackage{csquotes}

%% file: sections/section01-UG.tex
\section{Introduction} 

Computation of information theoretic quantities in quantum field theory has advanced significantly in the last two decades thanks to development of the replica trick \cite{EEandQFT_Calabrese_2004,EEandCFT_Calabrese_2009} and holographic formulation of entanglement entropy in AdS/CFT correspondence \cite{Ryu_2006,towards_Casini_2011,holographic_calculatins_Hung_2011}. Another method that helped obtain exact results was the development of supersymmetric localization techniques, which allows one to reduce path integrals in strongly interacting supersymmetric gauge theories to ordinary integrals over their Coulomb branch \cite{Pestun_2012,Kapustin_Yaakov_exact}.

In particular, Nishioka and Yaakov \cite{SUSYRenyi_Nishioka_2013} obtained the exact partition function of $3d$ $\mathcal{N}=2$ superconformal gauge theories on a branched three-sphere with a conical singularity along a great circle (denoted here by $S^3_n$) by means of supersymmetric localization, which allowed for the computation of flat space R\'enyi entropy across a circle. This follows from the standard replica trick, and Casini, Huerta and Myers' formulation of R\'enyi entropy of a sphere in flat space in terms of thermal partition functions on hyperbolic cylinders \cite{towards_Casini_2011}, and equivalently on $S^3_n$. The proposed definition of \textit{supersymmetric Rényi entropy} is then 
\begin{equation}\label{susyRenyi}
    S_n^{\text{susy}}=\frac{1}{1-n}\Re\left[\log\left(\frac{Z_{S_n^3}}{(Z_{S^3})^n}\right)\right]\text{,}
\end{equation}
while entanglement entropy is obtained as the limit $S_{\mathcal{A}}^{\text{susy}}=\lim_{n\to1}S_n^{\text{susy}}$. Interesting applications of this result include the duality between spacetimes with hyperbolic black holes and supersymmetric gauge theories \cite{qSCFTHuang_2014,gravdualNishioka:2014mwa},  entanglement entropy of a single quark in SYM \cite{quarkLewkowycz:2013laa} and additional checks of gauge-gravity duality in $4d$ $\mathcal{N}=4$ SYM \cite{Crossley:2014oea}. 

One feature of Nishioka and Yaakov's supersymmetric definition of R\'enyi entropy (\ref{susyRenyi}) is the introduction of a background $R$ symmetry gauge field that is necessary to preserve supersymmetry on the branched sphere. This renders the definition of R\'enyi entropy in $3d$ $\mathcal{N}=2$ superconformal theories inherently dependent on this deformation of the field content. The computation is also highly dependent on conformal symmetry of the theory and having a spherically symmetric entangling region to map the computation of flat space R\'enyi entropy to a partition function on a compact space where localization computations are well-established.

In an attempt to generalise our understanding of the entanglement structure of the theories under consideration, in this work we consider breaking the latter requirement and allow for smooth deformations of the entangling region around a circle. We make use of powerful results in the literature that constrain the geometry of three-manifolds on which supersymmetric theories can be consistently defined (in the paradigm of the rigid limit of supergravity theories \cite{Rigid_Festuccia_2011}) as well as their supersymmetry-preserving geometric deformations.

We start with a brief review of the necessary ingredients to compute (\ref{susyRenyi}), following \cite{SUSYRenyi_Nishioka_2013}. We then present our analysis of dependence of $S_{\mathcal{A}}^{\text{susy}}$ on deformations of the entangling surface. The main result of this section is that $S^{\text{susy}}_{\mathcal{A}}$ only captures non-local correlations which are blind to the specific geometry of the entangling region, depending on it only through its topology. Therefore, $S_{\mathcal{A}}^ {\text{susy}}$ does not compute a quantity with (all of) the usual properties associated with von Neumann entropy. Intrigued by this result, we consider an alternative method to compute $S_{\mathcal{A}}^{\text{susy}}$ to gain better insight into this quantity. This makes use of the heat kernel expansion; the details of our method are explained in the second part of the paper.

\section{Review of Supersymmetric R\'enyi Entropy}
\noindent\textbf{Generalities. }We are interested in entanglement entropy of the vacuum state of $2+1$-dimensional QFT's on flat spacetime $\mathcal{M}$ with \textit{entangling region} given by a two-dimensional disk (to be denoted by $\mathcal{A}$). This is computed by the \textit{von Neumann} entropy
\begin{equation}
    S(\rho_{\mathcal{A}})=-\text{Tr}_{\mathcal{A}}[\rho_{\mathcal{A}}\log\rho_{\mathcal{A}}]\, ,
\end{equation}
where $\rho_{\mathcal{A}}=\text{Tr}_{\overline{\mathcal{A}}}[\rho]$ is the reduced density matrix of subregion $\mathcal{A}$, with $\rho$ being the full density matrix. A useful one parameter generalization of $S(\rho_{\mathcal{A}})$ is the \textit{Rényi entropy}, given by
\begin{equation}
    S_n(\rho_{\mathcal{A}})=\frac{1}{1-n}\log[\text{Tr}_{\mathcal{A}}\rho_{\mathcal{A}}^n]\, .
\end{equation}
This reduces to the von Neumann entropy in the $n\to 1$ limit upon analytic continuation in $n$. By considering the Euclidean path integral representation of density matrices, the trace of the $n$-th power of $\rho_{\mathcal{A}}$ may be expressed in terms of a partition function on the background spacetime of the QFT with conical singularities along the boundary of the entangling region (denoted by $\mathcal{M}_n$) \cite{EEandQFT_Calabrese_2004}. The Rényi entropy becomes
\begin{equation}
    S_n(\rho_{\mathcal{A}})=\frac{1}{1-n}\log\left(\frac{Z_{\mathcal{M}_n}}{Z_{\mathcal{M}}^n}\right)\text{.}
\end{equation}
This is significantly more tractable than the von Neumann entropy. In the presence of conformal symmetry, $Z_{\mathcal{M}_n}$ is equal to a partition function on a branched three-sphere with the location of the conical singularity being mapped to a great circle along $S^3$. The reason is that there exists a conformal transformation which maps the causal development\footnote{When talking about causal development we of course take $\mathcal{A}$ to be a time slice of a Lorentzian manifold. However, in the calculations of the following section we will implicitly consider flat space in Euclidean signature.} of the disk $\mathcal{A}$ on $\mathcal{M}_n$ to $S_n^3$ \cite{towards_Casini_2011}, the latter having the following metric in Euclidean signature:
\begin{equation}\label{metric_Sn}   ds^2_{S_n^3}=d\theta^2+n^2\sin^2\theta d\tau^2+\cos^2\theta d\phi^2\text{,}
\end{equation}
where $\theta\in\left[0,\frac{\pi}{2}\right]$ and $\tau,\phi\in[0,2\pi)$. This reduces to the metric on $S^3$ for $n=1$ and leads to the formal definition of supersymmetric Rényi entropy (\ref{susyRenyi}) for arbitrary $n$.

In the presence of supersymmetry, the partition functions $Z_{S^3}$ and $Z_{S^3_n}$ can be computed \textit{exactly} using supersymmetric localization. This technique relies on the presence of a fermionic operator $\mathcal{Q}$ which generates a symmetry of some action $S$. A simple argument then shows \cite{Pestun_2012} that the following deformation of the partition function,
\begin{equation}
    Z(t)=\int\mathcal{D}\varphi e^{-S[\varphi]-t\mathcal{Q}V[\varphi]}\text{,}
\end{equation}
is independent of $t$. Here, $\varphi$ denotes all the fields in the theory, while $V$ is some functional of the fields satisfying $\mathcal{Q}^2V=0$. Independence of $Z(t)$ on $t$ implies that it can be computed in the $t\to +\infty$ limit, for which the only field configurations contributing to the path integral are $\varphi_0$ such that $\mathcal{Q}V[\varphi_0]=0$, as well as quadratic fluctuations around $\varphi_0$. 

Supersymmetric gauge theories possess fermionic symmetry generators such as the $\mathcal{Q}$ above. These are generated by supercharges parametrized by the Killing spinors (see below) of the background geometry $\mathcal{M}$. For this reason, it is necessary to analyse the supersymmetries that are preserved on $\mathcal{M}$ before proceding with localization. When performing localization on the (branched cover of) the three-sphere, $\mathcal{Q}$ will be a linear combination of two superchages of opposite $R$-charge preserved on this background. On the other hand, choosing a single Killing spinor suffices even if there is more than one preserved supercharge on $\mathcal{M}$. 

Moreover, it can be shown that a functional $V[\varphi]$ with aforementioned properties exists in $3d$ $\mathcal{N}=2$ super Yang-Mills theories, and the supersymmetric actions for gauge fields and charged matter fields can be written in the form $\mathcal{Q}V$ \cite{bdry_Sugishita_2013}. This means that the saddle point approximation around stationary field configurations of these actions will give the exact partition function, up to an overall integration over the \textit{localization orbits} represented by the set of fields $\varphi_0$ above. In $3d$ $\mathcal{N}=2$ theories on manifolds without boundaries, this set corresponds to constant, Lie algebra-valued modes for $\sigma$, one of the scalar fields in the vector multiplet transforming in the adjoint representation of the gauge group. The final result reads
\begin{equation}
    Z=\int [d\sigma_0]Z_{\text{matter}}^{\text{1-loop}}(\sigma_0)Z_{\text{gauge}}^{\text{1-loop}}(\sigma_0)e^{-S(\sigma_0)}\text{,}
\end{equation}
where $[d\sigma_0]$ denotes the integration measure on the Lie algebra of the gauge group and $Z^{\text{1-loop}}(\sigma_0)$ are the result of the quadratic path integrals of the matter and gauge Lagrangians.
\vspace{0.4cm}

\noindent\textbf{Supersymmetry on the branched sphere. }In order to define our SCFT's on $S^3_n$, we must analyse the necessary conditions for preserving supersymmetry on this background. When placing a supersymmetric theory on a curved manifold $\mathcal{M}$,  supersymmetry algebra receives corrections which vanish in the limit of infinite curvature radius. This allows us to preserve only a subset of the flat space supercharges. Such corrections will depend on fields belonging to the supergravity multiplet, which contain the following fields (see \cite{SUSYfieldtheories_Closset_2013})
\begin{equation}    \mathcal{H}\subset\left(g_{\mu\nu},\Psi^{\alpha},\tilde{\Psi}^{\dot{\alpha}},A_{\mu}^{(R)},V_{\mu},H\right)\text{.}
\end{equation}
These are set to constant background configurations which is achieved by sendind $M_{Planck}\to \infty$ \cite{Rigid_Festuccia_2011}. Preserved supersymmetries on $\mathcal{M}$ are then parametrized by spinors which solve the \textit{Killing spinor equation} (KSE).

First consider the round $S^3$. Solutions of $R$-charges $\pm 1$ can be obtained by simply turning on $H=\mp i$ and setting the remaining background fields to zero; this leads to the KSE's
\begin{equation}
    \nabla_{\mu}\varepsilon =\pm\frac{i}{2}\sigma_{\mu}\varepsilon
\end{equation}
The simplest solutions are obtained for a choice of vierbein given by the left- or right- invariant vector fields $e^{\mu}_i$ on $S^3$, see \cite{Kapustin_Yaakov_exact,SUSYRenyi_Nishioka_2013}. This is because the spin connection greatly simplifies in this basis, $\omega_{ij}=\pm\varepsilon_{ijk}e^k$. The KSE for solutions of positive $R$-charge can be solved by any constant spinor in the left-invariant basis by turning on $H=-i$:
\begin{equation}   \nabla_{\mu}\varepsilon= \left(\partial_{\mu}+\frac{1}{8}e^k_{\mu}\varepsilon_{ijk}[\sigma^i,\sigma^j]\right)\varepsilon=\left(\partial_{\mu}+\frac{i}{2}\sigma_{\mu}\right)\varepsilon
\end{equation}
This gives two linearly independent constant Killing spinors. Analgously, there are two other Killing spinors which solve the KSE of negative $R$-charge by considering now the right-invariant vierbein basis with $H=i$, for which
\begin{equation}    \nabla_{\mu}\varepsilon=\left(\partial_{\mu}-\frac{i}{2}\sigma_{\mu}\right)\varepsilon\text{.}
\end{equation}
This shows that $S^3$ preserves all four (real) supercharges of the $\mathcal{N}=2$ algebra.

Moving on to $S^3_n$, if one tries to solve the KSE with the same supergravity background, one finds two independent solutions which are not globally defined because of the conical singularity at $\theta=0$. This can be seen from the integrability condition implied by the KSE (and the analogous condition for the solution of $R$ charge $-1$, $\tilde{\zeta}$). One needs to cancel the conical singurarity without turning on singular background fields. There exists a judicious way out \cite{SUSYRenyi_Nishioka_2013} if one endows $\zeta$ and $\tilde{\zeta}$ with opposite and non-vanishing $R$-charge. This yields a non-trivial $A_{\mu}$, gauging the $R$ symmetry while retaining a vanishing central charge of the supersymmetry algebra, $i.e.$ $V_{\mu}=0$. The integrability condition in this case reads
\begin{equation}\label{integ_cond}
    \left[\frac{i}{2}(2R_{\mu\nu}-Rg_{\mu\nu})\gamma^{\nu}-2i\varepsilon_{\mu\nu\rho}\nabla^{\nu}A^{\rho}+iH^2\gamma_{\mu}\right]\zeta=0\, .   
\end{equation}
The singularity at $\theta=0$ due to presence of $\delta(\theta)$ in the Ricci scalar is cancelled by the field strength of $A$ if one choses \begin{equation}
    A=\frac{n-1}{2}\text{d}\tau\, .
\end{equation}
We emphasize that the quantity (\ref{susyRenyi}) can only be computed in the presence of an additional background field which gauges the $R$ symmetry. This is essential in order to preserve supersymmetry on the branched three-sphere, and therefore to perform localization.

Finally, a regularization of the background geometry and supergravity fields is necessary, because the  supergravity background above preserves too many supercharges for a theory with a singular $R$ symmetry field strength according to general conditions derived in \cite{SUSYfieldtheories_Closset_2013}. By redefining the line element as $g_{\theta\theta}=f_{\delta}^{-1}(\theta)$ and the background fields as
\begin{equation}\label{reg_sugra_back}
    H=-i\sqrt{f_{\epsilon}(\theta)}\text{,}\,\,\,\,\,\,A=\frac{q\sqrt{f_{\epsilon}(\theta)}-1}{2}\text{d}\tau\text{,}\,\,\,\,\,\,V=0\text{,}
\end{equation}
one can derive that two constant independent Killing spinors are preserved.

\section{Deformations of $\partial\mathcal{A}$}

\subsection{Trivial deformations of $S^3$ and entangling surfaces}\label{secIIIB}

Compact, oriented three manifolds $\mathcal{M}_3$ which preserve at least two supercharges admit a Killing vector. Indeed, given two Killing spinors of opposite $R$ charge, $\zeta$ and $\tilde{\zeta}$, we have
\begin{equation}    K^{\mu}=\zeta\gamma^{\mu}\tilde{\zeta}\Rightarrow\nabla_{(\mu}K_{\nu)}=0\text{,}
\end{equation}
by virtue of $\zeta,\tilde{\zeta}$ satisfying the Killing spinor equation. One can additionally define the following tensors on $\mathcal{M}_3$:
\begin{equation}
    \eta_{\mu}=\frac{1}{\Omega}K_{\mu}\text{,}\,\,\,\,\,\,\Phi^{\mu}_{\,\,\,\nu}=\varepsilon^{\mu}_{\,\,\,\nu\rho}\eta^{\rho}\text{,}
\end{equation}
where $\Omega^2=K_{\mu}K^{\mu}$ is a normalization factor such that $\eta_{\mu}\eta^{\mu}=1$. It is immediate to see that $\Phi$ satisfies
\begin{equation}\label{AMS}
    \Phi^{\mu}_{\,\,\,\rho}\Phi^{\rho}_{\,\,\,\nu}=-\delta^{\mu}_{\,\,\,\nu}+\eta^{\mu}\eta_{\nu}\text{.}
\end{equation}
This, together with the fact that the following integrability condition holds,
\begin{equation}\label{int_cond}
    \Phi^{\mu}_{\,\,\,\nu}(\mathcal{L}_K\Phi)^{\nu}_{\,\,\,\rho}=0\text{,}
\end{equation}
(where $\mathcal{L}_K$ is the Lie derivative along $K^{\mu}$) implies that the triple $(K^{\mu},\eta_{\mu},\Phi^{\mu}_{\,\,\,\nu})$ defines a \textit{tranversely holomorphic foliation} (THF) on $\mathcal{M}_3$. When $\mathcal{M}_3$ admits such a geometric structure, it can be shown \cite{SUSYfieldtheories_Closset_2013} that there exist adapted coordinates on $\mathcal{M}_3$ for which the metric takes the form
\begin{equation}\label{adapted_metric}
ds^2_{\mathcal{M}_3}=\left(d\psi+h(z,\bar{z})dz+\bar{h}(z,\bar{z})d\bar{z}\right)^2+c(z,\bar{z})^2dzd\bar{z}\text{.}
\end{equation}
It was shown in \cite{geom_Closset_2014,twisted_Closset:2014uda} that supersymmetric partition functions $Z_{\mathcal{M}_3}$ are independent of the parameters $c(z,\bar{z})$, $h(z,\bar{z})$ and $\bar{h}(z,\bar{z})$. 

Our strategy is to take these trivial deformations of the adapted metric on $S^3$ (and branched covers thereof) and conformally map them to flat space through the Casini-Huerta-Myers (CHM) map \cite{towards_Casini_2011}. This is a conformal transformation which maps the causal development of the disk on an equal (Euclidean) time slice of $\mathbb{R}^3$ to $S^3$, allowing for the equivalence between flat space Rényi entropy across a circle and (\ref{susyRenyi}). Under (the inverse of) this conformal transformation the line element on $S^3$ changes as 
\begin{equation}
ds^2_{S^3}\xrightarrow{\text{CHM}}\Omega^{-2}ds^2_{\mathbb{R}^3}\text{,}
\end{equation}
for some conformal factor $\Omega^2$ whose particular form is not relevant. Under the procedure described above, we will have
\begin{equation}\label{def_map} 
\begin{aligned}
&ds^2_{S^3}+\Delta_cds^2_{S^3}+\Delta_hds^2_{S^3}
\\&\hspace{1.6cm}\xrightarrow{\text{CHM}}\Omega^{-2}\left[ds^2_{\mathbb{R}^3}+\Delta_cds^2_{\mathbb{R}^3}+\Delta_hds^2_{\mathbb{R}^3}\right]
\end{aligned}
\end{equation}
The right hand side contains metric deformations around flat space which are conformally equivalent to deformations of $c(z,\bar{z})$ and $h(z,\bar{z})$ and do not affect supersymmetric partition functions, leaving the R\'enyi entropy invariant. Because we do not want to alter the background spacetime geometry of the entangling region, the next step is to determine which of these flat space deformations are pure gauge, realised by diffeomorphisms $\xi^{\mu}$ such that 
\begin{equation}    (\Delta_c+\Delta_h)g_{\mu\nu}=\nabla_{(\mu}\xi_{\nu)}\text{.}
\end{equation}
Evaluating such diffeomorphisms at $\rho=R$ determines trivial deformations of the boundary of the entangling surface that the flat space R\'enyi entropy invariant. 

Denoting an entangling region obtained by smooth deformations of the disk by $\tilde{\mathcal{\mathcal{A}}}$,  we want to find such $\tilde{\mathcal{A}}$ that
\begin{equation}
    Z_{\text{CHM}[\mathcal{D}_{\tilde{\mathcal{A}}}]}=Z_{S^3}\text{,}
\end{equation}
where $\text{CHM}[\mathcal{D}_{\tilde{\mathcal{A}}}]$ denotes the image of the causal development of $\tilde{A}$ under the CHM map (where it is understood that $\hat{A}$ is in Lorentzian signature).

An important aspect regarding the validity of this procedure is that the KMS property of the CHM
map, as shown to hold in \cite{towards_Casini_2011} for circular entangling surfaces, is still obeyed if we apply
it to a deformed entangling surface.
This is because it relies on the modular flow induced from the modular Hamiltonian
$H$, defined as $\rho=e^{-H}$, which always exists in a relativistic field theory and need not be local, allowing us to depart from circular entangling regions.

There is one final remark which has crucial consequences for the outcome of this section. It was shown in \cite{twisted_Closset:2014uda} the parameter independence of $Z_{\mathcal{M}_3}$ stated below (\ref{adapted_metric}) applies to deformations around an arbitrary THF (even though this independence was only explicitly derived for deformations around flat space in \cite{geom_Closset_2014}). This is shown by performing a so-called $R$-symmetry twist of the SUSY algebra, which renders the supercharge a scalar under transformations of adapted coordinates. Since $\delta_{\zeta}$ is a scalar, $\delta_{\zeta}$-exact terms in $\Delta\mathcal{L}_{\mathbb{R}^3}$ remain $\delta_{\zeta}$-exact in the Lagrangian variation around $\mathcal{L}_{\mathcal{M}_3}$ for any $\mathcal{M}_3$. A consequence of this is that $Z_{\mathcal{M}_3}$ is also independent of deformations of the THF around $(K+\Delta K,\eta+\Delta\eta,\Phi+\Delta\Phi)$, since this is a well-defined THF by virtue of $\Delta K$, $\Delta\eta$, $\Delta\Phi$ satisfying the required integrability conditions derived in \cite{geom_Closset_2014}. This means that $Z_{\mathcal{M}_3}$ remains invariant after applying an arbitrary number of infinitesimal deformations to the THF. The consequence is that we need not restrict to deformations of the parameters $c$ and $h$ with infinitesimal amplitude.
\vspace{0.4cm}

\noindent\textbf{Deformations of $c(z,\bar{z})$. }We consider the three-sphere $S^3$ parametrized in Hopf coordinates, for which the line element is (\ref{metric_Sn}) evaluated at $n=1$. The relation between these coordinates and the ones parametrizing the adapted metric (\ref{adapted_metric}) can be found in Appendix \ref{appendix1}. 

We consider a parametrization of flat Euclidean spacetime in polar coordinates $(t,\rho,\varphi)$ and the metric $ds^2_{S^3}$ deformed by the line element
\begin{equation}
\Delta cdzd\bar{z}=\Delta c\left(f^{\prime2}d\theta^{2}+f^{2}(d\varphi^{2}+d\tau^{2}-2d\varphi d\tau)\right)\text{,}
\end{equation}
where the function $f(\theta)$ is defined in Appendix \ref{appendix1}. We perform the sequence of CHM transformations, keeping
track of the conformal factors for the transformation between
$ds_{S^{3}}^{2}$ and the flat space metric as in (\ref{def_map}). This task simplifies significantly by restricting to the time slice $t=0$.\footnote{One might worry that this may no longer correspond
to a spacelike surface because we have now acquired a crossed term
$\sim d\varphi dt$ in the flat space metric, altering the norm
of the normal vector $\xi_{\mu}=(1,0,0)$ in the timelike direction.
However, by continuity of $\Delta c$ there will be no change in the sign of this
norm, and the $t=0$ hypersurface remains spacelike.} In the end, we obtain the following deformation of the $t=0$ spacelike surface in flat space:
\begin{equation}
\Delta_{c}ds^{2}=\Delta c^{2}\left[f^{\prime2}d\rho^{2}+f^{2}\frac{R^{2}}{4}\left(1+\frac{\rho^{2}}{R^{2}}\right)^{2}d\varphi^{2}\right]\text{.}
\end{equation}

Now recall that $\Delta c^{2}$ is arbitrary as
long as it is real. In particular, we
can give $\Delta c$ an arbitrary $\varphi$ dependence, which will
result in a non-symmetric profile for the entangling region around
the origin of the plane.

At $t=0$ the function $f(\theta)$ is expressed in terms of $\rho$ as (see Appendix \ref{appendix1})
\begin{equation}
f^2(\theta)\Big|_{t=0}=4\left(\frac{R^2-\rho^2}{R^2+\rho^2}\right)^2\text{,}
\end{equation}
and also
\begin{equation}
f^{\prime 2}(\theta)\Big|_{t=0}=\frac{16R^2\rho^2}{(R^2+\rho^2)^2}\text{.}
\end{equation}
This gives
\begin{equation}\label{deltacmetric}
\begin{aligned}
\Delta_{c}ds^{2}&=\Delta c^{2}\Bigg[\frac{16R^2\rho^2}{(R^2+\rho^2)^2}d\rho^{2}+\\
&\,\,\,\,\,\,\,\,\,\,+R^2\left(\frac{R^2-\rho^2}{R^2+\rho^2}\right)^2\left(1+\frac{\rho^{2}}{R^{2}}\right)^{2}d\varphi^{2}\Bigg]\text{.}
\end{aligned}
\end{equation}
\vspace{0.4cm}

\noindent\textbf{Trivial diffeomorphisms and integrability conditions. } As discussed above, we are interested in deformations induced by diffeomorphisms $x^{\mu}\to x^{\mu}+\xi^{\mu}(x)$ 
such that
\begin{equation}\label{diffeo}
\Delta_{c}g_{\mu\nu}=2\nabla_{(\mu}\xi_{\nu)}\text{.}
\end{equation}
From this condition we obtain the following integrability conditions:
\begin{equation}\label{diff_integ_cond}
\begin{cases}
\partial_{\rho}\xi_{\rho}(\rho,\varphi)=\Delta c^{2}\frac{16R^2\rho^2}{(R^2+\rho^2)^2}\\
\partial_{\varphi}\xi_{\varphi}(\rho,\varphi)=\Delta c^{2}R^2\left(\frac{R^2-\rho^2}{R^2+\rho^2}\right)^2\left(1+\frac{\rho^{2}}{R^{2}}\right)^{2}\\
\xi_{t}(\rho,\varphi)=0\text{,}
\end{cases}
\end{equation}
where we ruled out timelike translations of the entangling surface, and omitted time dependence of $\xi^{\mu}$ as we work on a fixed time slice $t=0$. 

Finally, we analyse how these diffeomorphisms act on the boundary of the entangling surface. We then solve the integrability conditions (\ref{diff_integ_cond}) at $\rho=R$ of the $t=0$ surface. The second condition
is trivially solved as long as $\Delta c^{2}$ is an integrable
function of $\varphi$:
\begin{equation}
 \xi^{\varphi}(\rho=R,\varphi)=k(\rho=R)\text{,}
\end{equation}
where $k(\rho)$ is an arbitrary function of $\rho$; the undeformed metric was used to map forms to vectors.
Because of the symmetry of the entangling surface along $\varphi$,
the action of this diffeomorphism along the $\varphi$ direction is trivial and leaves
the entangling surface invariant.

We may also restrict to the case where $\Delta c^{2}$ has no $\rho$
dependence (such a dependence would only contribute with some constant rescaling
of $\xi^{\rho}$, given that we will evaluate it at fixed
$\rho$). This way, the first condition can be integrated to give
\begin{equation}
\xi^{\rho}(\rho=R,\varphi) =2(\pi-2)R\Delta c^2(\varphi)\text{,}
\end{equation}
where we ignored any residual $t$ or $\rho$ dependence from integration
constants. Note in particular that $\xi^{\rho}$ is directly proportional
to $R$. Deformations of $h(z,\bar{z})$ can be analysed in a similar manner; the resulting diffeomorphisms lead to the same conclusions as above.

The same conclusions can be derived with respect to deformations around the branched sphere by retracing the same steps. The subtelty here is that both the metric and the background fields need to be regularized to obtain globally defined and regular solutions to Killing spinor equations as well as to the integrability conditions which determine the supersymmetry-preserving deformations of the THF. This  regularization was achieved in \cite{SUSYRenyi_Nishioka_2013}.

We have obtained a diffeomorphism acting on the boundary of the entangling region whose radial component has an arbitrary angular dependence. This implements arbitrary smooth deformations of $\partial\mathcal{A}$. Because these diffeomorphisms result from deformations of the THF on $S^3$ and $S^3_b$ which do not affect the partition functions $Z_{S^3}$ and $Z_{S^3_b}$ (again, as long as we talk about partition functions on regularized squashed sphere backgrounds), we conclude that \emph{vacuum entanglement entropy of $3d$ $\mathcal{N}=2$ theories is independent of arbitrary smooth deformations of the boundary of the circular entangling region on flat space.} It is worth emphasizing that, due to independence of $Z_{\mathcal{M}_3}$ on $h(z,\bar{z})$ and $c(z,\bar{z})$, our conclusion holds also for  deformations of $\partial\mathcal{A}$ with finite amplitude. This result is not completely novel as it was implicit in the literature and known to authors of \cite{SUSYRenyi_Nishioka_2013}, \footnote{We acknowledge our correspondence with I. Yaakov on this point.} but we nevertheless find it very useful to present an explicit derivation as it emphasizes the topological feature of the supersymmetric Renyi entropy defined in \cite{SUSYRenyi_Nishioka_2013} in contrast to the usual von Neumann entropy. 

\subsection{Squashed spheres and corrections to $S_{\mathcal{A}}^{\text{susy}}$}\label{sec_IIIB}

Having found that a large class of diffeomorphisms preserve the disk entanglement entropy, one can now ask which  deformations of $S^3$ affect the partition function, leading to deformations of entanglement across deformed circles. The first part of this question was answered by \cite{geom_Closset_2014}; such deformations turn out to be realized by $\textit{squashed spheres}$, whose metric reads
\begin{equation}\label{squashed_metric}
ds^{2}=d\theta^{2}+b^{2}\sin^{2}\theta d\tau^{2}+b^{-2}\cos^{2}\theta d\varphi^{2}\text{,}
\end{equation}
for some $b\in\mathbb{R}$ and coordinate ranges as in \ref{metric_Sn}. We know that this metric possesses a conical singularity with deficit
angle $2\pi(1-b^{2})$ around $\theta=0$ and $(1-b^{-2})2\pi$ around $\theta=\frac{\pi}{2}$. These are mapped by CHM to conical singularities on flat space at $\rho=R$ and $\rho=0$, respectively. The latter is produced by a twisting of the $t=0$ timeslice of flat space along the angular direction $\varphi$.
\vspace{0.4cm}

\noindent\textbf{Non-trivial deformations of the THF on $S^3$. }The non-trivial deformations of the THF are parametrized by the cohomology
classes of certain $(0,1)$-forms in the cohomology group $H^1(S^3)$ with coefficients
in the holomorphic tangent bundle $T^{1,0}\mathcal{M}_{3}$.
It is shown in \cite{geom_Closset_2014} that there are only two distinct cohomology classes, given
by
\begin{equation}\label{non-trivial-cl}
\Theta^{(\gamma)}=\gamma X\otimes\eta\text{,}\,\,\,\,\,X=\begin{cases}
z(\partial_{z}-h\partial_{\psi})\\
(\partial_{z}-h\partial_{\psi})\text{ or }z^{2}(\partial_{z}-h\partial_{\psi})\text{.}
\end{cases}
\end{equation}
All the known examples of squashed spheres preserving two supercharges are realized by deformations
of the first type above, and these are the ones we will focus on.\footnote{In the second type, the moduli $\gamma$ can in fact be absorbed by a rescaling of the holomorphic coordinate on $\mathbb{CP}^1$, since under $z\to\gamma z$ or $z\to\gamma^{-1}z$, the vector fields $(\partial_z-h\partial_{\psi})$ and $z^2(\partial_z-h\partial_{\psi})$, respectively, are rescaled by a factor of $\gamma^{-1}$. For this reason, the second type does not seem to correspond to a physically relevant deformation (at least in a conformal theory). This is because we expect that such deformations can be recast in terms of partial derivatives of the free energy of the theory with respect to the moduli, which is equivalent to the insertion of some integrated operators; we will see how this works below for first type. We thank I. Yaakov for pointing this out to us.}

Knowing the relevant class of deformations of $S^3$ which $Z_{S^3}$ depends on, we can determine the corresponding deformations around flat space as was done for the trivial deformations. For this we choose to work with the metric and the one-form $\Theta^{(\gamma)}$ in coordinates adapted to the THF.

For the deformation of interest, the holomorphic component of $\Theta^{(\gamma)}$ is
\begin{equation}
\Theta^{z}=\gamma z(d\psi+hdz+\bar{h}d\bar{z})\text{.}
\end{equation}
Comparing this with the generic expression for $\Theta^z$ in terms of the deformation parameters of the THF (section $5.3$ of \cite{geom_Closset_2014}), we identify
\begin{equation}
\begin{cases}
\Delta K^{z}=\frac{i}{2}\gamma z\\
\Delta\Phi_{\,\,\,\bar{z}}^{z}=-\frac{1}{2}\gamma z\bar{h}\text{.}
\end{cases}
\end{equation}
Note that, by construction, these deformations satisfy the required integrability conditions \cite{geom_Closset_2014}. These parameters
correspond to an explicit deformation of the Killing vector $K^{\mu}$ which
determines the orbits of the Seifert fibration:
\begin{equation}
\begin{aligned} & K=\partial_{\psi}\longrightarrow K+\Delta K=\partial_{\psi}+\frac{i}{2}\gamma z\partial_{z}\text{.}\end{aligned}
\end{equation}
In turn, this induces a deformation of the dual one-form $\eta_{\mu}$ and, consequently,
of the adapted metric. To first order in the deformations, we have
\begin{equation}
\begin{aligned}\eta_{\mu} & =(g_{\mu\nu}+\Delta g_{\mu\nu})(K^{\nu}+\Delta K^{\nu})\Rightarrow\\
\Rightarrow\Delta\eta & =g_{\mu\nu}\Delta K^{\nu}dx^{\mu}+\Delta g_{\mu\nu}K^{\nu}dx^{\mu}\\
 & =\frac{i}{2}\gamma z\left((c+2|h|^{2})d\bar{z}+h^{2}dz+2\bar{h}d\psi\right)+\Delta g_{\mu\psi}dx^{\mu}\text{.}
\end{aligned}
\end{equation}

Because $Z_{\mathcal{M}_3}$ is independent
of $\Delta K^{\bar{z}}$ (or depends holomorphically on $\Delta K$),
we can set $\Delta K^{\bar{z}}=0$. The same applies to $\Delta\Phi_{\,\,\,z}^{\bar{z}}$ and its complex conjugate $\Delta\Phi_{\,\,\,\bar{z}}^{z}$.
\footnote{Such a choice does not affect the reality of the partition function: $\Delta g_{\psi\psi}$
written below will be real, because so is the combination $ihz=\frac{i}{2}(ig_{2}(\theta)e^{-i\phi})f(\theta)e^{i\phi}$.}

The requirement that $g_{\mu\nu}+\Delta g_{\mu\nu}$
remains transversely Hermitean leads to \cite{geom_Closset_2014}
\begin{equation}\label{type}
\begin{cases}
\Delta g_{\psi\psi}=-i\gamma hz\\
\Delta g_{\psi z}=\frac{i}{2}\gamma zh^{2}+\Delta g_{\psi z}-\frac{i}{2}\gamma h^{2}z\\
\Delta g_{zz}=i\gamma zh^{3}+\Delta g_{\psi z}\text{.}
\end{cases}
\end{equation}
The second condition is satisfied trivially for any $\Delta g_{\psi z}$.
This is a reflection of the fact that the parameter $\Delta\eta_{z}$
is a moduli of the geometry, since $Z_{\mathcal{M}_3}$ does
not depend on it. This is equivalent to independence of $Z_{\mathcal{M}_3}$ on $h(z,\bar{z})$. Therefore $\Delta g_{\psi z}$
can be set to zero or tuned to set $\Delta g_{zz}=0$.

We conclude that the moduli deformation $\Theta^{(\gamma)}$
leads to a variation of the adapted metric which always includes a
non-zero component $\Delta g_{\psi\psi}=-i\gamma hz$.
Applying the CHM map, the resulting line element at $t=0$ is
\begin{equation}
\begin{aligned}-i\gamma hzd\psi^{2}=\frac{\gamma R^{2}}{16}\bigg(1-2\left(\frac{R^{2}-\rho^{2}}{R^{2}+\rho^{2}}\right)^{2}\bigg)\left(1+\frac{\rho^{2}}{R^{2}}\right)^{2}d\varphi^{2}\end{aligned}
\text{.}
\end{equation}
At $\rho=R$, this gives precisely the line element of a circle with
a defect angle proportional to $\gamma$. The integrability condition of the corresponding diffeomorphisms is now
\begin{equation}
\partial_{\varphi}\xi_{\varphi}(t,\rho,\varphi)=\frac{\gamma R^{2}}{16}\left(1-2\left(\frac{R^{2}-\rho^{2}}{R^{2}+\rho^{2}}\right)^{2}\right)\left(1+\frac{\rho^{2}}{R^{2}}\right)^{2}\text{,}
\end{equation}
which is trivially solved as
\begin{equation}
    \xi^{\varphi}(\rho=R,\varphi)=\frac{\gamma}{4}\varphi\text{.}
\end{equation}

This (local) diffeomorphism rotates the boundary of the entangling surface along itself in the angular direction, so it does not alter its shape. However, because $\varphi$ is an angular coordinate identified as $\varphi\sim\varphi + 2\pi$, $\xi^{\varphi}$ is not continuous at $\varphi=0$ and therefore does not represent a smooth deformation of $\partial\mathcal{A}$. This discontinuity signals the presence of a conical singularity at $\rho=0$ and tells us that $\xi^{\varphi}$ is "twisting" the circular entangling surface over itself. This was expected due to the conical singularity of the standard squashed sphere metric (\ref{squashed_metric}) at $\theta=\frac{\pi}{2}$.

This further supports our conclusion in the previous section. Indeed, we find that the only instance in which small deformations of the circular entangling surface affect the vacuum entanglement is when these correspond to inserting a conical singularity at the centre of the disk. In this case, the topology of $\mathcal{A}$ is no longer that of a disk, while the entanglement entropy remains invariant under infinitesimal smooth deformations of its boundary.
\vspace{0.4cm}

\noindent\textbf{Corrections to $S_{\mathcal{A}}^{\text{susy}}$ under a conical deformation. }We will now compute the correction to the entanglement entropy under the circular non-trivial deformation of the entangling surface with the conical defect at $\rho=0$. The knowledge of this correction will completely determine the entanglement strucutre of our SCFT's for infinitesimal deformations of $\mathcal{A}$.

Instead of working with the Lagrangian variation $\Delta\mathcal{L}$ written in terms of
adapted coordinates on the Seifert manifold, we can take a more direct
approach and consider a particular squashing of $S^3$ in Hopf coordinates considered in \cite{SUSYfieldtheories_Closset_2013}, where most of the relevant computations have been done, which we review below. The metric is
\begin{equation}\label{squashedclossetmetric}
ds^{2}=\frac{r^{2}}{4}\left(h^{2}(d\psi+2\sin^{2}\frac{\theta}{2}d\phi)^{2}+(d\theta^{2}+\sin^{2}\theta d\phi^{2})\right)\text{,}
\end{equation}
with $h\equiv\frac{1}{2}(b+b^{-1})$, and $\theta\in[0,\pi]$, $\psi,\phi\in[0,2\pi)$.

It is possible to determine  the supergravity background exactly, since
this squashed sphere falls into the category of three-manifolds for
which the nowhere vanishing Killing vector $V^{\mu}$ determines a
fibration over a surface of constant curvature, either $S^{2},T^{2}$
or $H^{2}$; see \cite{SUSYfieldtheories_Closset_2013} for details. 

We express (\ref{squashedclossetmetric}) in coordinates which reduce to stereographic coordinates on $S^3$ when $b=1$:
\begin{equation}
g_{\mu\nu}=\Omega^{2}\delta_{\mu\nu}+\left(\frac{b-b^{-1}}{b+b^{-1}}\right)^{2}v_{\mu}v_{\nu}\text{,}\,\,\,\,\,\,\,\,\,\Omega=\frac{2}{1+x^{2}}\text{.}
\end{equation}
Here, $v_{\mu}$ relates to the nowhere vanishing Killing vector $V_{\mu}$ which determines the fibration of (\ref{squashedclossetmetric}) over an $S^2$ (see section 8 of \cite{SUSYfieldtheories_Closset_2013} for the explicit expression for $v_{\mu}$).

Before proceeding, we should make sure that this corresponds to the first-type deformation of THF in (\ref{type}), since, as emphasised in \cite{geom_Closset_2014}, there exist squashings which do not change the THF but only the adapted metric, and therefore would leave the partition function invariant. In \cite{geom_Closset_2014} a two-parameter squashed sphere is constructed, and the corresponding THF is seen to be realised by a deformation of the first type around the three-sphere. There, the case $\gamma_i=0$ corresponds to the squashed sphere with $SU(2)\times U(1)$ isometry presented in (\ref{squashedclossetmetric}).\footnote{This can be seen by directly subsituting $\gamma_i=0$, from which one obtains the line element in the form $ds^2=\eta^2+2g_{z\bar{z}}dzd\bar{z}$, with
\[
\eta=\frac{2}{4+\gamma_r^2}\left(-\cos^2\frac{\theta}{2}d\phi+\sin^2\frac{\theta}{2}d\psi\right)\text{,}\,\,\,\,\,\,\,2g_{z\bar{z}}=\sin\theta(d\phi+d\tau)
\]
}

Another way to see this is the following: consider the linearized variation
of the Lagrangian to $\mathcal{O}(\delta b)$ ($\delta b=b-1$). Because the first order term in $\delta b$ of the expansion of $g_{\mu\nu}$ is zero, at $\mathcal{O}(\delta b)$ we have
\begin{equation}
\begin{aligned}
\delta\mathcal{L}^{(\delta b)}&=\left(A^{\mu}-\frac{3}{2}V^{\mu}\right)j_{\mu}^{(R)}+j_{\mu}^{(Z)}C^{\mu}\\
&=\delta bv^{\mu}(ij_{\mu}^{(Z)}-j_{\mu}^{(R)})\text{.}
\end{aligned}
\end{equation}
This cannot correspond to a $Q$-exact Lagrangian variation, because
neither of the $Q$-exact bosonic operators which we can form with
our superalgebra ((6.3) of \cite{geom_Closset_2014}) contain components of $j_{\mu}^{(R)}$,
only of its derivatives.

We emphasize that the energy-momentum tensor is not included in $\delta\mathcal{L}$
because the leading order is $\delta b^{2}$. In the case of a superconformal theory, the operator $j_{\mu}^{(Z)}$
belongs to the supermultiplet containing $T_{\mu}^{\mu}$,
and is therefore redundant. Morevover, one-point functions of operators in a CFT vanish in general. The lowest order correction to the partition
function is then
\begin{equation}
\begin{aligned}
Z^{(\delta b)}&=\int\mathcal{D}\varphi e^{-S-\int d^{3}x\sqrt{g}\delta\mathcal{L}^{(\delta b)}}\\
&=Z^{(0)}\left(1+\delta b^2\mathcal{I}_{S^3}\right)\text{,}
\end{aligned}
\end{equation}
where
\begin{equation}\label{I_S3}
    \mathcal{I}_{S^3}=\frac{1}{2}\int d^{3}x\sqrt{g}\int d^{3}y\sqrt{g}v_{\mu}(x)v_{\nu}(y)\langle j^{(R)\mu}(x)j^{(R)\nu}(y)\rangle_{S^{3}}
\end{equation}
Regularizing divergent contributions from the contact terms, one can show that \cite{SUSYfieldtheories_Closset_2013}
\begin{equation}
    \mathcal{I}_{S^3}=-\frac{1}{2}\frac{\partial^2 F}{\partial b^2}\Bigg|_{b=1}=-\frac{\pi^2}{4}\tau_{rr}+\text{(imaginary terms)}\text{,}
\end{equation}
where the imaginary terms originate from background Chern-Simons terms which must be included to regularize divergent contributions from contact terms while giving an imaginary contribution to the free energy, see \cite{contact_Closset_2012,comments_Closset:2012vp}; $\tau_{rr}>0$ is the coefficient of the flat space two point function of the $R$ symmetry current at separated points (we set $\tau_{rr}=1$ below). 

We now want to study the integrated correlator (\ref{I_S3}) on $S^3_n$. We claim that (\ref{I_S3}) only depends on $n$ through the integration measure. Indeed, the form of $\langle j^{(R)\mu}(x)j^{(R)\nu}(x)\rangle_{S^3}$ is fixed by conformal symmetry and any dependence on $n$ can be absorbed in $x$, with the integration range changing accordingly; by the same argument, $v_{\mu}(x)$ should not depend on $n$. We moreover assume that $\tau_{rr}$ does not depend on $n$ since, as observed in \cite{comments_Closset:2012vp}, this should be determined by the field content of the theory (for a fixed energy scale). In our framework, Rényi entropies are computed keeping the field content fixed while branching the background geometry. 

The above implies that
\begin{equation}\label{no_cont}  
\mathcal{I}_{S^3_n}=n^2\mathcal{I}_{S^3}\,\,\,\,\Rightarrow \,\,\,\,\partial_n\mathcal{I}_{S_n^3}|_{n=1}=2\mathcal{I}_{S_3}\text{.}
\end{equation}
To obtain the correction to the entanglement entropy, we proceed as follows:
\begin{equation}\label{Sb}
\begin{aligned}
    \delta_b S_{\mathcal{A}}&=\lim_{n\to 1} \frac{1}{1-n}\log\left(\frac{Z_{S^3_n}^{(\delta b)}}{(Z_{S^3}^{(\delta b)})^n}\right)\\
    & = \lim_{n\to 1} \frac{1}{1-n}\bigg[(1-n)\log(1+(\delta b)^2\mathcal{I}_{S^3_n}|_{n=1})+\\
    &\,\,\,\,\,+(n-1)\partial_n\log(1+(\delta b)^2\mathcal{I}_{S^3_n})|_{n=1}+\mathcal{O}((1-n)^2)\bigg]\\
    & = \log(1+(\delta b)^2\mathcal{I}_{S^3})-\frac{2(\delta b)^2\mathcal{I}_{S^3}}{1+(\delta b)^2\mathcal{I}_{S^3}} \\
    &=-(\delta b)^2\mathcal{I}_{S^3}+\mathcal{O}(\delta b^4)\text{.}
\end{aligned}
\end{equation}

\begin{figure}
    \centering    \includegraphics[scale=0.55]{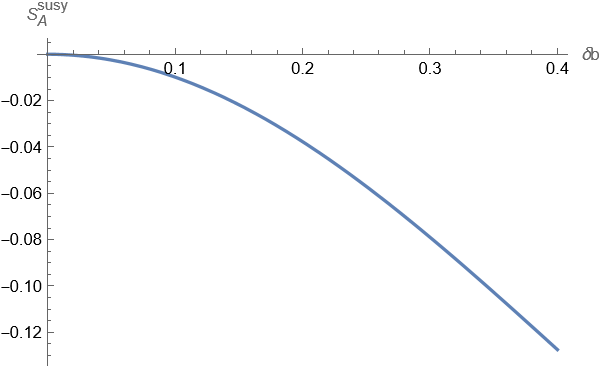}
    \caption{Correction to supersymmetric entanglement entropy as a function of the excess angle of the conical defect introduced at the center of $\mathcal{A}$.}
    \label{fig:deltab}
\end{figure}

As shown in Figure \ref{fig:deltab}, the behaviour of this lowest order correction to $S_{\mathcal{A}}^{\text{susy}}$ is simple: it is a concave function of the excess angle introduced by the conical singularity.

In \cite{disksmaxBueno:2021fxb} it has been shown that the universal term of the entanglement entropy is maximized by circles in $3d$ among all regions, including ones with non-trivial topologies with  $n_B>1$ smooth boundaries.\footnote{In PM's Master's thesis we have gathered some evidence that the same observation regarding regions with $n_B>1$ boundaries holds in our $3d$ $\mathcal{N}=2$ SCFT's of interest.} However, topologies with singular regions such as the one in the present section were not considered; we have shown that the same behaviour holds for such regions, thus extending the scope of the conclusions of \cite{disksmaxBueno:2021fxb} for regions with conical topology.

This result has relied on the knowledge of (a perturbative expansion of) the partition function of our SCFT's on a \textit{branched cover of the squashed sphere}. It was shown in \cite{qSCFTHuang_2014} that the exact result for such a partition function equals that of a branched sphere, which implies that the Rényi entropy on the branched squashed sphere in the $n\to 1$ limit equals minus the free energy on the branched sphere. This is a locally convex function, \textit{i.e} $\partial_b^2|_{b=1}F>0$, so we see that the local behaviour of the entanglement entropy around the disk topology obtained from the leading order perturbative result agrees with the global behaviour expected from the non-perturbative result.
\section{Entanglement Entropy and the Heat Kernel}
In this section we study entanglement entropy of $3d$ ${\cal N}=2$ gauge theories from a different angle, by employing the heat kernel expansion \cite{Birrell_Davies}. The heat kernel method has been applied to study entanglement structure of QFTs successfully in the past, see e.g. \cite{EE_Nishioka_2018, Vassilevich:2003xt, Solodukhin:2011gn} for a collection of results.   

First, consider a non-interacting quantum field theory defined on $\mathcal{M}_n$, an $n$-branched cover of a smooth Riemannian manifold $\mathcal{M}$ with a conical singularity along a given codimension-two surface $\Sigma$. If the quadratic operator for a given field is of the form $\mathcal{D}+m^{2}$, with $\mathcal{D}$ some differential operator, the one-loop effective action on $\mathcal{M}_n$ is \cite{EE_Nishioka_2018}
\begin{equation}\label{kernel}
\log Z_{n}=-\frac{1}{2}\log\det(\mathcal{D}+m^{2})=\frac{1}{2}\int_{\epsilon^{2}}^{\infty}\frac{ds}{s}\text{tr}\,K_{\mathcal{M}_{n}}(s)e^{-m^{2}s}\text{,}
\end{equation}
and the heat kernel operator has the following expansion:
\begin{equation}
\text{tr}\, K_{\mathcal{M}_{n}}(s)=\text{tr}\, e^{-s\mathcal{D}}=\frac{1}{(4\pi s)^{d/2}}\sum_{i=0}^{\infty}a_{i}(\mathcal{M}_{n})s^{i}\text{.}
\end{equation}

Heat kernel coefficients $a_i(\mathcal{M}_n)$ allow for expansion of one-loop effective action in the $n\to1$
limit, from which entanglement entropy can be obtained. One has 
\begin{equation}\label{heat_coeff}
a_{i}=a_{i}^{\text{bulk}}+(1-n)a_{i}^{\Sigma}+\mathcal{O}((1-n)^{2})\text{,}
\end{equation}
where $a_i^{\text{bulk}}$ do not include any contribution from the singularity and are given by an integral over the smooth part of $\mathcal{M}_n$; therefore $a_i^{\text{bulk}}(\mathcal{M}_n)=na_i^{\text{bulk}}(\mathcal{M})$. The second term $a_{i}^{\Sigma}$ represents  contributions from the conical singularity on the spacelike hypersurface
$\Sigma$ and contains non-trivial information to compute the entanglement
entropy. We will additionally need to consider a branched manifold with regularized
conical singularities, $\tilde{\mathcal{M}}_{n}$. By computing $a_{i}^{\text{bulk}}(\tilde{\mathcal{M}}_{n})$,
the surface parts $a_{i}^{\Sigma}$ can be extracted by reading off
its coefficient in $(1-n)$. This is because $a_i^{\text{bulk}}(\tilde{\mathcal{M}}_n)$ is smooth over the entire (regulated) manifold, with the singularities captured by the terms that vanish in the $n\to 1$ limit. It will include terms in $(1-n)$ which diverge when removing the regularization and contain the non-trivial contributions from the conical singularity.

Upon performing the $s$ integral in (\ref{kernel}), the final form for the entanglement entropy, explicitly
dependent on the UV cutoff, is
\begin{equation}
\begin{aligned}
S_{\mathcal{A}}&=\lim_{n\to 1}\frac{1}{1-n}\left(\log Z_n-n\log Z_1\right)\\
&=\frac{1}{2(4\pi)^{d/2}}\sum_{i=0}^{\infty}\frac{a_{i}^{\Sigma}}{m^{2i-d}}\Gamma\left(i-\frac{d}{2},m^{2}\epsilon^{2}\right)\text{,}
\end{aligned}
\end{equation}
where $\Gamma(a,x)$ is the incomplete Gamma function. 
From an expansion of the incomplete Gamma
function around $\epsilon=0$, we have
\begin{equation}
\begin{aligned}\Gamma\left(-\frac{3}{2},m^{2}\epsilon^{2}\right) & =\frac{2}{3m^{3}\epsilon^{3}}+\frac{2}{m\epsilon}+\mathcal{O}(\epsilon)\\
\Gamma\left(-\frac{1}{2},m^{2}\epsilon^{2}\right) & =\frac{2}{m\epsilon}+\mathcal{O}(\epsilon)\text{.}
\end{aligned}
\end{equation}
We find that the entropy in $3d$ can be expressed as 
\begin{equation}
\begin{aligned}
S_{\mathcal{A}}&=\frac{1}{2(4\pi)^{3/2}}\left[a_{0}^{\Sigma}\left(\frac{2}{3\epsilon^{3}}+\frac{2}{m\epsilon}\right)+a_{1}^{\Sigma}\frac{2}{\epsilon}\right]+\mathcal{O}(\epsilon)\\ &=\frac{1}{(4\pi)^{3/2}}\frac{a_{1}^{\Sigma}}{\epsilon}\text{.}
\end{aligned}
\end{equation}
We learn that in $3d$ we are only interested in $a_{1}^{\Sigma}$ (it turns out $a_{0}^{\Sigma}$ vanishes, see below).

\subsection{Heat kernel coefficients of our SCFT's and the zero coupling limit}\label{sec7.2}

We will collect the bulk and surface parts of the heat kernel coefficients for the
quadratic operators which compute the one-loop determinants after
localization on $S_{n}^{3}$, where $\Sigma$ is
the entangling surface at $\theta=0$. The relevant heat kernel coefficients will correspond the Lagrangians for the free theories describing the UV limit of our IR $\mathcal{N}=2$ SCFT's. This identification is made by virtue of localization: when taking the $t\to\infty$ limit in the localization procedure we are in the weak Yang-Mills coupling limit (because in SYM we identify $t\equiv g_{YM}^{-1}$), which corresponds to the UV due to asymptotic freedom.

We use the regularized supergravity background as well as the regularized branched sphere metric (see (\ref{reg_sugra_back}) above).
Upon localizing our SCFT we have a free theory with one scalar and
one fermionic field (the remaining scalar field from the chiral multiplet is
non-dynamical and can trivially be integrated out).
\vspace{0.4cm}

\noindent\textbf{Scalar heat kernel coefficients. }The scalar quadratic
operator in the matter Lagrangian is (choosing the superconformal $R$-charge $r=\frac{1}{2}$ and ommiting the dependence on $l$, the radius of the $S^3$)
\begin{equation}\label{D_phi}
\begin{aligned} &-\nabla^{2}+\frac{1}{4}A_{\mu}A^{\mu}+\left(\frac{1}{2}H+\sigma_{0}\right)^{2}-H^{2}-\frac{1}{8}(R-6H^{2})\\
 =& -\nabla^{2}+\frac{1}{n^{2}\sin^{2}\theta}\frac{(n\sqrt{f_{\delta}(\theta)}-1)^{2}}{16}-i\sigma_{0}+\sigma_{0}^{2}-\frac{1}{8}R\text{,}
\end{aligned}
\end{equation}

We can use the results from \cite{distributional_Fursaev_2013} for the surface heat cofficients
for a scalar quadratic operator of the form $\mathcal{D}=-\nabla^{2}+V$, where we identify $V$ with the terms added to the Laplacian in (\ref{D_phi}). These are
\begin{equation}\label{surf_coeff}
\begin{aligned} & a_{0}^{\Sigma}=0\text{,}\,\,\,\,\,\,\,\,\,\,a_{1}^{\Sigma}=\frac{2\pi}{3}(1-6\xi)\int_{\Sigma}1\text{,}\end{aligned}
\end{equation}
where $\xi$ is the coupling of $\phi$ to the Ricci scalar.

Note that the above expression for $a_{1}^{\Sigma}$ can be recovered from the one for the bulk coefficients presented in \cite{cones_Fursaev_1997}:
\begin{equation}\label{bulk_coeff}
a_{1}^{\text{bulk}}(\mathcal{M}_{n})=n\int_{\mathcal{M}}\sqrt{g}d^{d}x\left(\frac{1}{6}R-V\right)\text{.}
\end{equation}
The factor $\frac{2\pi}{3}\int_{\Sigma}1$ from $a_1^{\Sigma}$ in (\ref{surf_coeff}) is exactly the $(1-n)$
contribution coming from the term $\frac{1}{6}R$ in this equation, when it is integrated over the regularized $S^3_n$. This can be checked explicitly
in our case, making use of the Ricci scalar of the regularized $S^3_n$ presented in Appendix C of \cite{SUSYRenyi_Nishioka_2013}. 
Because $(\mathcal{D}+m^2)_{\phi}$ contains $-\frac{1}{8}R$ in the potential, the singularities in the Ricci scalar give the following overall contribution to the surface coefficients:
\begin{equation}\label{surf_scalar}
a_{\phi}^{\Sigma}=\left(1+\frac{6}{8}\right)\frac{2\pi}{3}\int_{\Sigma}1=\frac{7\pi}{6}\int_{\Sigma}1\text{.}
\end{equation}

It is shown in Appendix \ref{appendix2} that the potential terms on the branched sphere do not contribute to the surface part of the heat kernel coefficient. The contribution of the scalar field to the entanglement
entropy is
\begin{equation}\label{ent_scalar}
\frac{1}{(4\pi)^{3/2}}\frac{a_{1,\phi}^{\Sigma}}{\epsilon}=\frac{7\sqrt{\pi}}{24}\frac{l}{\epsilon}\text{.}
\end{equation}

\noindent\textbf{Spinor heat kernel  coefficients. } The generalization of (\ref{bulk_coeff}) to fields of higher spin reads \cite{cones_Fursaev_1997}
\begin{equation}
a_{1,(j)}^{\text{bulk}}=\int_{\mathcal{M}}\left(\frac{N^{(j)}}{6}R\text{Tr}\textbf{1}-\text{Tr}_{i}X^{(j)}\right)\text{,}
\end{equation}
where $N^{(j)}$ is the dimension of the representation of the Lorentz group of the spin $j$  field, $\text{Tr}_i$ is trace over indices of the representation, and $X^{(j)}$ is a potential. With the superconformal $R$-charge $r=\frac{1}{2}$, the quadratic operator for the spinor fields in the matter Lagrangian reads
\begin{equation}
  -i\slashed{\nabla}-i\sigma_{0}-\frac{1}{2}\gamma^{\mu}A_{\mu}\text{,}
\end{equation}
where we use the conventions of \cite{SUSYRenyi_Nishioka_2013} for the $\gamma$-matrices. We again refer to Appendix \ref{appendix2} for a proof of the vanishing contribution from potential terms. The surface coefficient for spin $\frac{1}{2}$ fields is then found in a similar way to $a_{1,\phi}^{\Sigma}$ by (inserting $\xi=\frac{1}{4}$)
\begin{equation}
a_{1,\psi}^{\Sigma}=-\frac{2^{\lfloor 3/2\rfloor}}{2}\left(1-\frac{6}{4}\right)\frac{2\pi}{3}\int_{\Sigma}1=\frac{2\pi^2l}{3}\text{,}
\end{equation}
so that the contribution from the fermions to the entanglement entropy is
\begin{equation}
    \frac{1}{(4\pi)^{3/2}}\frac{a_{1,\psi}^{\Sigma}}{\epsilon}=\frac{\sqrt{\pi}}{12}\frac{l}{\epsilon}\text{.}
\end{equation}

\noindent\textbf{Vector heat kernel coefficients. }After gauge fixing the path integral and integrating out ghost fields and the quadratic fluctuations in $\sigma$, the quadratic operator for the fields in the vector multiplet is 
\cite{SUSYRenyi_Nishioka_2013}
\begin{widetext}
\begin{equation}\label{gauge_L}
\begin{aligned}\mathcal{L}_{\text{gauge}} & =\text{Tr}\left[B_{\mu}\Delta_{v}B^{\mu}-[B_{\mu},\sigma_{0}]^{2}-i\tilde{\lambda}\gamma^{\mu}(\nabla_{\mu}+iA_{\mu})\lambda-i\tilde{\lambda}[\sigma_{0},\lambda]+\frac{1}{2}\tilde{\lambda}\lambda\right] =\sum_{i=1}^{2}\text{Tr}\left[B_{\mu}^{i}\Delta_{v}B_{i}^{\mu}-i\tilde{\lambda}_{i}\gamma^{\mu}(\nabla_{\mu}+iA_{\mu}-\frac{i}{2})\lambda_{i}\right]\\
 & \,\,\,\,\,\,\,\,\,\,+\sum_{\alpha\in \Phi}\text{Tr}\left[B_{\mu}^{-\alpha}(\Delta_{v}+\alpha(\sigma_{0})^{2})B_{\alpha}^{\mu}-i\tilde{\lambda}_{-\alpha}\gamma^{\mu}(\nabla_{\mu}+iA_{\mu})\lambda_{\alpha}+\tilde{\lambda}_{-\alpha}(-i\alpha(\sigma_{0})+\frac{1}{2})\lambda_{\alpha}\right]\text{.}
\end{aligned}
\end{equation}
\end{widetext}
Here $\Delta_{v}=\star d\star d+d\star d\star$ and $B_{\mu}$ is
the divergenceless part of the photon field
. 
Moreover, the fields from the gauge multiplet have been decomposed in the Cartan basis of $\mathfrak{g}$ as well as in the root space $\Phi$ of
the adjoint representation, where the labels $-\alpha$ are to be understood as minus the vector $\alpha$ in the vector space $\mathfrak{g}$. 

The gaugino quadratic determinant gives the same heat kernel coefficients
as the fermions in the chiral multiplet, and the surface parts vanish
for the same reasons as above; this means $a^{\Sigma}_{1,\lambda}=a^{\Sigma}_{1,\psi}$. It remains to include the heat kernel coefficients for the vector field
\begin{equation}
a_{1,B}^{\Sigma}=N^{(1)}a_{1,\phi}^{\Sigma}=\frac{7\pi}{3}\int_{\Sigma}1\text{,}
\end{equation}
where $N^{(1)}=2$ since the divergenceless vector $B_{\mu}$ has
two off-shell degrees of freedom. In the end we still need to multiply the coefficients
of the photon and gauginos by the dimension of the Cartan subalgebra
plus the number of root subspaces $\alpha$ of $\mathfrak{g}$, which
we denote by $N_{\text{Cartan}}+N_{\alpha}$. In \cite{cones_Fursaev_1997} there is an additional
term $-4\pi\int_{\Sigma}1$, but this comes from the presence of $R_{\mu\nu}$
in the vector Laplacian, which is not present in our case; its contribution
is precisely the $(1-n)$ term in $-\int_{\tilde{S}_{n}^{3}}\text{Tr}R_{\mu\nu}=-\int_{\tilde{S}_{n}^{3}}g^{\mu\nu}R_{\mu\nu}=-4\pi\int_{\Sigma}$,
so we remove this.

Putting everything together, we
have the following UV divergent contributions in the entanglement entropy of the matter and
gauge Lagrangians at zero coupling:
\begin{equation}
\begin{aligned}S_{\mathcal{A},\Phi}^{(\epsilon)} & =\frac{1}{(4\pi)^{3/2}}\frac{1}{\epsilon}(a_{1,\phi}^{\Sigma}+a_{1,\psi}^{\Sigma})=\frac{3\sqrt{\pi}}{8}\frac{l}{\epsilon}\text{,}\\
S_{\mathcal{A},\mathcal{V}}^{(\epsilon)} & =\frac{1}{(4\pi)^{3/2}}\frac{1}{\epsilon}(N_{\text{Cartan}}+N_{\alpha})(a_{1,\lambda}^{\Sigma}+a_{1,B}^{\Sigma})\\&=\frac{35\sqrt{\pi}}{48}(N_{\text{Cartan}}+N_{\alpha})\frac{l}{\epsilon}\text{.}
\end{aligned}
\end{equation}

\subsection{Finite gauge coupling: handling the integration over $\sigma_{0}$}\label{sec7.3}

Our result above is valid only at zero coupling, when the partition function is a product of one-loop determinants for the matter and gauge Lagrangians. This is only in the case of an abelian gauge theory, where $\sigma_0$ does not appear in $\mathcal{L}_{YM}$. Here we consider the case when the matter content of the theory is coupled to gauge fields. The main difference is that the partition function no longer factorizes into a product of one-loop determinants, but there is the overall integration over the localization orbits,

\begin{equation}
Z_{n}=\int[d\sigma_{0}]e^{-S_{\text{cl}}(\sigma_{0},n)}Z_{\text{matter}}^{\text{1-loop}}(\sigma_{0},n)Z_{\text{gauge}}^{\text{1-loop}}(\sigma_{0},n)\text{.}
\end{equation}

In this situation we cannot compute $\log Z_n$ by simply taking the logarithm of the full expression and retaining the surface heat kernel coefficients, due to the overall integration over the Lie algebra $\mathfrak{g}$. However, we can still make use
of the heat kernel expansion for each individual $1$-loop determinant
by rewriting the above as
\begin{equation}
\begin{aligned}
Z_{n}=&\int[d\sigma_{0}]e^{-S_{\text{cl}}(\sigma_{0},n)}\exp\left(\log Z_{\text{matter}}^{\text{1-loop}}(\sigma_{0},n)\right)\times\\
&\hspace{1.5cm}\times\exp\left(\log Z_{\text{gauge}}^{\text{1-loop}}(\sigma_{0},n)\right)\text{.}
\end{aligned}
\end{equation}
The price to pay is that we must consider the full expansion (\ref{heat_coeff})
incuding the bulk coefficients, since their contribution doesn not necessarily vanish when computing the Rényi entropy. However, it remains sufficient to consider
its expansion up to first order in $(1-n)$ if we are only interested in the entanglement entropy.
\vspace{0.4cm}

\noindent\textbf{Examples: Abelian gauge groups and topological sectors.}
\vspace{0.2cm}

\noindent\textit{Single $U(1)$ chiral multiplet. }Let us first consider a $U(1)$ gauge group, where the $\sigma_{0}$ dependence
drops out of $\mathcal{L}_{\text{gauge}}$ (\ref{gauge_L}) as the commutators  vanish. According to the discussion above, we compute the bulk heat coefficients for the scalar and spinor Lagrangians from the chiral multiplet. This is done in Appendix \ref{appendix4}, where we resolve a subtlety in defining the heat coefficients for the spinor Lagrangian, essentially due to its dependence on $\sigma_0$ which acts as a mass term. Here we note that the solution relies on a complexification of the variable of integration of the gauge moduli, and therefore all $\sigma_0$ terms present in the scalar bulk coefficient must also be complexified for consistency with this change of variable

The resulting expression for $\log Z_{\text{matter}}^{\text{1-loop}}$ is:
\begin{equation}
  n\frac{\sqrt{\pi}l^{2}}{4}\left(-\frac{5}{4l^{2}}+2(\sigma_{0}-\frac{1}{4l})^{2}-\frac{1}{8l^{2}}\right)\frac{l}{\epsilon}+(1-n)S^{
 (\epsilon)}_{\mathcal{A},\Phi}
\end{equation}
Since this is all the $\sigma_{0}$ dependence that is present in the UV partition function, we can perform the integral over the
Lie algebra. For the integral to converge, we go back to the original real contour through $\sigma_{0}\to-i\sigma_{0}$. We find
\begin{equation}
\int_{-\infty}^{+\infty}d\sigma_{0}e^{-n\sqrt{\pi}\frac{l^3}{2\epsilon}(\sigma_{0}-\frac{i}{4l})^{2}}=\frac{1}{l}\sqrt{\frac{2\sqrt{\pi}\epsilon}{nl}}\text{.}
\end{equation}

Expanding the Rényi entropy in the limit $n\to 1$, 

\begin{widetext}
    \begin{equation}
\begin{aligned}
&\frac{1}{1-n}\left[\log\int_{-\infty}^{+\infty}[d\sigma_0]\exp\left(\log Z_{\text{matter}}^{\text{1-loop}}(\sigma_{0},n)\right)-n\log\int_{-\infty}^{+\infty}[d\sigma_0]\exp\left(\log Z_{\text{matter}}^{\text{1-loop}}(\sigma_{0},1)\right)+(\text{gauge})\right]=\\
 =& \frac{1}{1-n}\left[(1-n)\left(\frac{3\sqrt{\pi}}{8}\frac{l}{\epsilon}-\frac{1}{2}\log\frac{l}{2\sqrt{\pi}\epsilon}+\frac{1}{2}+(S_{\mathcal{A},\Phi}^{(\epsilon)}+S^{(\epsilon)}_{\mathcal{A},\mathcal{V}}-\log 2\pi)\right)+\mathcal{O}((1-n)^{2})\right]\text{,}
\end{aligned}
\end{equation}
\end{widetext}
where we have recognised that the contribution from the gauge multiplet does not depend on $\sigma_0$, and therefore it amounts to $S_{\mathcal{A},\mathcal{V}}^{(\epsilon)}$. The terms inside
the exponential proportional to $n$ not contributing to the $\sigma_{0}$
integral will vanish in the expression for the entropy. This results in (with $S_{\mathcal{A},\mathcal{V}}^{(\epsilon)}$
computed with $N_{\alpha}=0$, $N_{\text{Cartan}}=1$)
\begin{equation}
\begin{aligned}S_{\mathcal{A}}^{U(1)} & =S_{\mathcal{A},\Phi}^{(\epsilon)}+S_{\mathcal{A},\mathcal{V}}^{(\epsilon)}-\frac{1}{2}\log\frac{l}{2\sqrt{\pi}\epsilon}-\log 2\pi+\frac{1}{2}\\
 & =\frac{7\sqrt{\pi}}{6}\frac{l}{\epsilon}-\frac{1}{2}\log\frac{l}{\epsilon}-\frac{1}{2}\log2-\frac{3}{4}\log\pi+\frac{1}{2}\text{.}
\end{aligned}
\end{equation}
We note that this computation is unchanged if we take the limit of infinite radius, $l\to +\infty$. In particular,

\begin{equation}
\begin{aligned}
   & \exp\left(\log Z_{\text{matter}}^{\text{1-loop}}(\sigma_{0},n)\right)_{l\to+\infty}=\\=&
   \exp\left(n\frac{\text{Vol}(S^3)}{(4\pi)^{3/2}}\frac{1}{\epsilon}2\sigma_0^2+(1-n)(S_{\mathcal{A},\Phi}^{(\epsilon)}+S_{\mathcal{A},\mathcal{V}}^{(\epsilon)})\right)\text{.}
\end{aligned}
\end{equation}
This leads to the same $\sigma_0$-integral and to the same final expression for $S_{\mathcal{A}}^{U(1)}$. Therefore, the only effects produced by taking $l\to+\infty$ are the IR divergences in the logarithmic term and in the linear term in $\epsilon^{-1}$.
This is a consequence of conformal symmetry: changing macroscopic scale of the geometry $l$ has no direct physical consequence and can be re-expressed in terms of a change in the UV cutoff (recall that $l$ is related to the radius of the circle $R$ in flat space, but can always be absorbed by a Weyl transformation of the metric in the conformal theory).

Before attempting to make sense of this result, let us work out several more examples with distinct gauge groups in order to study the general structure of the terms resulting from this computation.
\vspace{0.4cm}

\noindent\textit{Product of $N$
$U(1)$ chiral multiplets.} In this case the partition function simply factorizes into a product of $N$ integrals
over $\sigma_{0}^{(1)},...,\sigma_{0}^{(N)}$, each contributing the same amount
and the entropy is simply $N$ times the previous result:

\begin{equation}
S_{\mathcal{A}}^{U(1)^N}=N\left[\frac{7\sqrt{\pi}}{6}\frac{l}{\epsilon}-\frac{1}{2}\log\frac{l}{\epsilon}-\frac{1}{2}\log2-\frac{3}{4}\log\pi+\frac{1}{2}\right]\text{.}
\end{equation}
\vspace{0.4cm}

\noindent\textit{Two chiral multiplets in $\textbf{1}$ and $\overline{\textbf{1}}$
of $U(1)$.} Going through the same steps as in the single $U(1)$ chiral multiplet, $\exp\left(\log Z_{\text{matter}}^{\text{1-loop}}(\sigma_{0},n)\right)$ becomes
\begin{equation}
\exp\left(\frac{n\sqrt{\pi}}{4}\left(-\frac{5}{2}+4l^{2}\sigma_{0}^{2}\right)\frac{l}{\epsilon}+2(1-n)\frac{3\sqrt{\pi}}{8}\frac{l}{\epsilon}\right)\text{.}
\end{equation}
The integral over $\sigma_{0}$ is now
\begin{equation}
\int_{-\infty}^{+\infty}\frac{d\sigma_0}{\text{Vol}(U(1))} e^{-n\sqrt{\pi}\frac{l^{3}}{\epsilon}(\sigma_{0}-\frac{i}{4l})^{2}}=\frac{1}{2\pi l}\sqrt{\frac{\sqrt{\pi}\epsilon}{nl}}\text{,}
\end{equation}
which results in\footnote{Below we denote by $N_f$ the number of flavours of the theory, which we consider as the number of pairs of chiral multiplets in the fundamental and anti-fundamental representations of the guage group. $N$ denotes the number of colors of the theory, \textit{i.e.} the fields are charged under $U(N)$ and transform under representations of rank $N$.}
\begin{equation}
S_{\mathcal{A}}^{N=1,N_f=1}=2S^{(\epsilon)}_{\mathcal{A},\Phi}+S^{(\epsilon)}_{\mathcal{A},\mathcal{V}}-\frac{1}{2}\log\frac{l}{\epsilon}-\log 2-\frac{3}{4}\log\pi+\frac{1}{2}\text{.}
\end{equation}
This represents a difference of $-\frac{1}{2}\log2$
relative to the finite part of the previous case. Note that the coefficient
of $S_{\mathcal{A},\mathcal{V}}$ is not doubled because we still
have only one abelian gauge multiplet with $N_{\text{Cartan}}=1$
and $N_{\alpha}=0$. More interestingly, the coefficient of the logarithmic
divergence has remained the same as for the case of a single chiral $U(1)$
multiplet. This suggest that
this coefficient only depends on the rank of the gauge group. This
would be a natural conclusion, since the logarithmic divergence only
occurs when gauge fields are present at non-zero coupling with the
matter fields.
\vspace{0.4cm}

\noindent\textit{$U(1)$ chiral multiplet $+$ $U(1)_k$ Chern-Simons. }The integral over $\mathfrak{u}(1)$ is modified if we include a Chern-Simons
action, which introduces a term dependent on the classical action in the integrand, $e^{-S_{\text{cl}}(\sigma_{0},n)}=e^{in\pi k\sigma_{0}^{2}}$. Defining $a\equiv \frac{\text{Vol}(S^{3})}{l(4\pi)^{3/2}}=\frac{\sqrt{\pi}l^2}{4}$,
it reads
\begin{equation}
\begin{aligned} 
& \int_{-\infty}^{+\infty}\frac{(-i)d\sigma_{0}}{\text{Vol}(U(1))}\exp\left(n\left(2a\frac{l}{\epsilon}-i\pi l^2k\right)\sigma_{0}^{2}-na\frac{\sigma_0}{l}\frac{l}{\epsilon}\right)\text{.}
\end{aligned}
\end{equation}
 The difference relative to the above case lies in
the imaginary term in the denominator of the square root which results from the gaussian integration above; this adds
to the entropy
\begin{equation}
 -\frac{1}{4}\log\left(\left(\frac{1}{2\sqrt{\pi}}\frac{l}{\epsilon}\right)^{2}+k^{2}\right)-\log 2\pi+\frac{1}{2}+\ldots\text{,}
\end{equation}
where the ellipses denote imaginary terms which we discard.
We learn that the Chern-Simons sector does not affect the entropy
in the presence of a $U(1)$ chiral multiplet (SQED) in the UV limit
$\epsilon\to0$.
\vspace{0.4cm}

\noindent\textit{Chiral multiplet in the bifundamental $\textbf{1}\times\overline{\textbf{1}}$
representation of $U(1)_{k}\times U(1)_{-k}$. }The weight of the $\textbf{1}\times\overline{\textbf{1}}$ representation
of $U(1)_{k}\times U(1)_{-k}$ which enters in the Lagrangian is $\sigma-\tilde{\sigma}$
, with $\sigma,\tilde{\sigma}\in\mathbb{R}$. The classical partition
function is now $e^{-S_{\text{cl}}(\sigma_{0},n)}=e^{in\pi k(\sigma^{2}-\tilde{\sigma}^{2})}$.
With the change of integration variables $\sigma_{\pm}=\sigma\pm\tilde{\sigma}$, we have 
\begin{equation}  \int_{-\infty}^{+\infty}\frac{d\sigma_{+}d\sigma_{-}}{2\text{Vol}(U(1))}\exp\left(in\pi k\sigma_{+}\sigma_{-}+\left(na\frac{\sigma_{-}}{l}-2na\sigma_{-}^{2}\right)\frac{l}{\epsilon}\right)\text{.}
\end{equation}
With this gauge group we obtain a finite contribution
to the topological entanglement entropy coming from the Chern-Simons sector,
which is
\begin{equation}
\begin{aligned}
&\lim_{n\to 1}\frac{1}{1-n}\left[-\log n\pi k+n\log\pi k\right]=\\=&-\log k-2\log 2-2\log\pi+1\text{.}
\end{aligned}
\end{equation}
The first term precisely matches the Chern-Simons contribution for this gauge group encountered in \cite{SUSYRenyi_Nishioka_2013}.
\vspace{0.4cm}

\noindent\textit{Comments.} Let us scrutinize our findings. We have not obtained the expected additive contribution of $-\frac{1}{2}\log k$ for the $U(1)_k$ chiral multiplet while for the $U(1)_k\times U(1)_{-k}$ bifundamental multiplet we indeed obtain such a contribution, however the divergent piece is not retained in the final result. 

One can also worry that if a logarithmic divergence is present in our final result, then any finite contribution can always be absorbed by shifting the UV cutoff $\epsilon$ rendering it ambigious. However, a cutoff-independent quantity is given by the \textit{difference}, at fixed $\epsilon$, of the finite terms observed between theories with different numbers of matter and gauge multiplets and/or different gauge groups. This effectively allows us to deduce the universal contribution to the entanglement when varying these parameters of the theory.

We can safely conclude that, \emph{the contribution of each chiral multiplet,  charged under $U(1)$, is $-\frac{1}{2}\log 2$, in agreement with the examples studied in \cite{SUSYRenyi_Nishioka_2013} i.e. free chiral multiplet and the $U(1)_k\times U(1)_{-k}$ ABJM theory. Moreover, we are able to extract the amount of universal entanglement due to each $U(1)$ gauge multiplet, namely $-\frac{3}{4}\log \pi+\frac{1}{2}\approx -0.36$.} 
From the first three examples above we observe that the logarithmic divergence resulting from our calculations depends exclusively on the number of charge generators, or the rank of the gauge group, and not on the number of matter multiplets. Indeed, when computing the entropy of two chiral multiplets with opposite electric charge, this term remains the same as in the case of a single $U(1)$ chiral multiplet due to the fact that the number of charge generators is unchanged (and only the coefficient of $S_{\mathcal{A},\Phi}$ doubles). 
\vspace{0.4cm}

\noindent\textbf{Examples: Non-abelian unitary groups.}
\vspace{0.4cm}

\noindent\textit{$U(2)$ gauge group.} As a warm up we consider the gauge group $U(2)$. We take the Cartan subalgebra as the set of diagonal matrices $\sigma_{0}=\text{diag}(\sigma_{1},\sigma_{2})$,
$\sigma_{1},\sigma_{2}\in\mathbb{R}$. The roots and weights of $\mathfrak{u}(2)$
are 
\begin{equation}
\begin{aligned} & \rho_{i}(\sigma_{0})=\sigma_{i}\text{,}\,\,\,\,i\in\{1,2\}\text{,}\\
 & \alpha_{12}(\sigma_{0})=\sigma_{1}-\sigma_{2}=-\alpha_{21}(\sigma_{0})\text{,}
\end{aligned}
\end{equation}
so that the measure on the space of localization zero modes is
\begin{equation}
[d\sigma]=\frac{d\sigma_{1}d\sigma_{2}}{\text{Vol}(U(2))}\prod_{\alpha>0}\alpha(\sigma_{0})^{2}=\frac{d\sigma_{1}d\sigma_{2}}{\text{Vol}(U(2))}(\sigma_{1}-\sigma_{2})^{2}\,,
\end{equation}
where $\text{Vol}(U(2)) = (2\pi)^3$.
Apart from this change in integration measure, the computation goes through essentially as in the abelian case; one simply needs to take care to include the contributions in the Lagrangians from the full set of roots and weights. This is demonstrated in some detail in Appendix \ref{appendix4}, and in this subsection we will mainly restrict to reporting the final results. The final result for a $U(2)$ chiral multiplet is (again neglecting a factor proportional to $\log l$):
\begin{equation}\label{U(2)_chiral}
\begin{aligned} 
S_{\mathcal{A}}^{U(2)}=S_{\mathcal{A},\Phi}^{(\epsilon)}+4S_{\mathcal{A},\mathcal{V}}^{(\epsilon)}-2\log\frac{l}{\epsilon}+3\log 2-\frac{3}{2}\log 5+2\text{.}
\end{aligned}
\end{equation}
Note that the coefficient of the logarithmic
divergence is $-2=-\frac{4}{2}$ which agrees with the general form 
 $-\frac{\text{dim}\mathfrak{g}}{2}\log\frac{l}{\epsilon}$ derived in \cite{SS_Casini_2020}.

Analogously to what was observed in the abelian case, considering a chiral
multiplet charged under the gauge group $U(2)^{N}$ (or equivalently, $N$ chiral multiplets charged under independent representations of $U(2)$) results in $S_{\mathcal{A}}^{U(2)^N}=NS_{\mathcal{A}}^{U(2)}$.
\vspace{0.4cm}

For two chiral multiplets in the \textbf{2} and $\overline{\textbf{2}}$ representations of $U(2)$, we obtain 
\begin{equation}\label{2U(2)_chirals}
S_{\mathcal{A}}^{N=2,N_f=1}=2S_{\mathcal{A},\Phi}^{(\epsilon)}+4S_{\mathcal{A},\mathcal{V}}^{(\epsilon)}-2\log\frac{l}{\epsilon}+\log 2-\frac{3}{2}\log 3+2
\end{equation}

It is now certainly possible to read off the amount of universal entanglement coming from each $U(2)$ chiral multiplet as in the abelian case, as well as the one for each vector multiplet by considering gauge group $U(2)^N$. For instance, the former equals $-2\log 2-\frac{3}{2}\log\frac{3}{5}\approx -0.62<-\frac{1}{2}$, signalling that a $U(2)$ chiral multiplet introduces higher universal entanglement in the vacuum state compared to a $U(1)$ chiral multiplet; this should likely be attributed to its non-abelian statistics. 
\vspace{0.4cm}

\noindent\textit{Generalization: $U(N)$ gauge group with $N_f$ flavours. }The integral (\ref{integral}) can easily be computed for gauge group $U(N)$, for arbitrary $N$, through a simple change of variables (see Appendix \ref{appendix4}). The result is
\begin{equation}
\begin{aligned}
&S_{\mathcal{A}}^{U(N)}=S_{\mathcal{A},\Phi}^{(\epsilon)}+N(N-1)S_{\mathcal{A},\mathcal{V}}^{(\epsilon)}-\log(\text{Vol}(U(N)))+\\
&+\frac{N(N-1)}{2}\left[3\log 2-2\log\frac{l}{\epsilon}+2+\log \frac{(N-1)^2}{(4N-3)^{\frac{3}{2}}}\right]
\text{.}
\end{aligned}
\end{equation}

We also note that our expression has the following behaviour in the large $N$ limit:
\begin{equation}
    N^2\left[-2\log\frac{l}{\epsilon}+\frac{1}{4}+\frac{5}{2}\log N-\log\pi\right]\text{,}
\end{equation}
where the asymptotic expansion of the Barnes function was used and only terms of $\mathcal{O}(N^2)$ or $\mathcal{O}(N^2\log N)$ were kept.

We finally note the expression for the entanglement entropy of $N_f$ flavours of chiral multiplets charged under $U(N)$, which follows from applying (\ref{U(2)integral}) for a flavour of $U(2)$ in the general integral (\ref{U(N)integral}):
\begin{widetext}
\begin{equation}
\begin{aligned}
&S_{\mathcal{A}}^{U(N),N_f}=\\&=S_{\mathcal{A},\Phi}^{(\epsilon)}+N(N-1)S_{\mathcal{A},\mathcal{V}}^{(\epsilon)}-\log(\text{Vol}(U(N)))+\frac{N(N-1)}{2}\Bigg[(3-N_f)\log 2-2\log\frac{l}{\epsilon}+2+\log \frac{(N-1)^2}{(4(N-1)+N_f)^{\frac{3}{2}}}\Bigg]\text{.}
\end{aligned}
\end{equation}
\end{widetext}

\subsection{Superselection sectors and the origin of the logarithmic divergence}
A generic but puzzling aspect of the results in the previous section is the presence of a logarithmic divergence which is not expected from geometric considerations in three dimensions. In particular it is unclear in odd dimensions how to write down a dimensionless coefficient for this term in terms of curvature invariants. 

In the presence of global charges, however, \cite{SS_Casini_2020} offers a way out. When the underlying theory preserves a global symmetry group $G$, observables of the theory are given by the \emph{orbifolded} operator algebra $\mathcal{O}=\mathcal{F}/G$  where $\mathcal{F}$ is the whole algebra. In particular the Hilbert space of $\mathcal{F}$ factorizes into \emph{superselection sectors} of a given charge 
\begin{equation}
\mathcal{H}_{\mathcal{F}}=\bigoplus_{r,i}\mathcal{H}_{r,i}\text{.}
\end{equation}
where $r$ labels an irreducible representation and $i$ labels the states in that irrep.


In particular, based on the presence of superselection sectors, \cite{SS_Casini_2020} derives upper and lower bounds for the following \textit{difference} of mutual informations:
\begin{equation}\label{DeltaI}
    \Delta I\equiv I_{\mathcal{F}}(R_{1},R_{2})-I_{\mathcal{O}}(R_{1},R_{2})\text{,}
\end{equation}
where $I_{\mathcal{F}}(R_{1},R_{2})$ and $I_{\mathcal{O}}(R_{1},R_{2})$ are the mutual informations of two disjoint spheres $R_1$ and $R_2$ measured in the additive parts of the algebras $\mathcal{F}$ and $\mathcal{O}$. $\Delta I$ is a useful order parameter sensitive to presence of superselection sectors because $\mathcal{O}_{\text{add}}(R_1\cup R_2)$ does not contain intertwiners \cite{SS_Casini_2020}
supported on $R_1\cup R_2$, while $\mathcal{F}_{\text{add}}$ does.

For $U(1)$ symmetry group, these lower/upper bounds are estimated in the limiting case where $R_2$ approaches the complement of the ball $R_1$ with separation $\epsilon$:
\begin{equation}
    \Delta I\simeq\frac{1}{2}\log\frac{A}{\epsilon^{d-2}}\, ,
\end{equation}
where $A$ is the area of the entangling surface. 
It was argued in \cite{SS_Casini_2020} that such a contribution should come from charge-anticharge fluctuations across the boundary of the entangling region, with both charges separated by the surface $\partial R_1$. 
Moreover, it can seen to be proportional to $\log\langle Q^2\rangle$ with the previous argument leading to the conclusion $\langle Q^2\rangle\simeq A/\epsilon^{d-2}$.

For non-abelian $G$, \cite{SS_Casini_2020} also argues as to why, in the limit of large charge fluctuations across the boundary, $\langle Q_1^2\rangle>>1$, the charge fluctuations generated by $\mathcal{G}_i$, $i=1,...,\text{dim}(\mathfrak{g})$ 
should behave as those generated by $\text{dim}(\mathfrak{g})$ abelian generators; in other words, non-commutative effects become negligeble when $\epsilon\to 0$ and when charge fluctuations across the boundary become relevant.
Then, up to a subleading constant, one finds
\begin{equation}\label{deltaI}
    \Delta I\simeq\frac{d-2}{2}\text{dim}(\mathfrak{g})\log\frac{R}{\epsilon}\text{,}
\end{equation}
with $R$ being the radius of $R_1$.

Even though the argument we presented above provides a plausible explanation for appearence of logarithms, we still need to clarify why the analysis of \cite{SS_Casini_2020} applies. First, symmetries of the $\mathcal{N}=2$ theories we consider here are gauged instead of global. However, they effectively become global in the localization computation thanks to the condition $a_{\mu}=0$  imposed on the localization orbits.\footnote{Note that this condition holds after gauge-fixing and the ghosts integrated out.} As a result, the local gauge symmetry of the IR theory freezes and \cite{SS_Casini_2020} applies. Second, for \cite{SS_Casini_2020} to apply, we must be computing entanglement entropy measured by an algebra of local operators in the orbifold theory. This is also true, effectively, as we compute entanglement in the vacuum state in which only uncharged operators have non-vanishing vacuum expectation value.

To be more explicit, the vacuum reduced
density matrix of $A$ should have zero eigenvalues on the Hilbert subspace corresponding to charged operators smeared inside $A$, which belong to non-trivial superselection sectors. Therefore, $S(\rho)$ should only pick up
contributions either from uncharged operators smeared in $A$ acting on $|0\rangle$ or from
intertwiners whose global charge is zero and hence do not annihilate $|0\rangle$; rather, they are given by the product of charged operators localized in $\mathcal{A}$ and its complement. The latter
leads to the logarithmic contribution, as shown in this section, and their presence can therefore be traced to the existence of intertwiners. 



\section{Discussion and Outlook. }
\noindent\textbf{$S_{\mathcal{A}}^{\text{susy}}$ is topological. } Our conclusion in section \ref{secIIIB} seems to contradict \cite{def_Mezei_2015} stating that spherical entangling surfaces locally minimize the universal term in the entanglement entropy of a CFT. There, it was shown that for infinitesimal smooth deformations of $\partial\mathcal{A}$ around a spherical entangling surface (parametrized by $\epsilon$), the universal piece of the CFT entanglement entropy in $d$ dimensions takes the form
\begin{equation}\label{locally_min}  S(\partial\mathcal{A})=S_d^{\text{sphere}}+\epsilon^2C_TF[\partial\mathcal{A}]+\mathcal{O}(\epsilon^3)\text{,}
\end{equation}
where $C_T$, determined by the stress tensor two-point function, and $F[\partial\mathcal{A}]$, a functional of the shape deformation, are strictly positive coefficients. This result from holography was  confirmed in \cite{Faulkner:2015csl}  using CFT techniques.
Given this, our conclusions related to the trivial deformations of $\partial\mathcal{A}$ establish that $F[\partial\mathcal{A}]$ vanishes for $3d$ $\mathcal{N}=2$ superconformal theories. 

In fact this does not contradict the cited results because what we consider is the supersymmetric entanglement entropy whose definition involves deformation of the theory by a background configuration for the $R$ symmetry gauge field $A_{\mu}^{(R)}$ in order to preserve supersymmetry in the presence of conical singularities. Therefore $S_{n}^{\text{susy}}$ departs from the usual definition of von Neumann entropy. 
We believe this difference is responsible for the deformation invariance we observe above.  
Our results suggest that \emph{background field deformation of $3d$ $\mathcal{N}=2$ superconformal theories required to define supersymmetric R\'enyi entropy renders it topological.} This statement applies only to  dependence of $S^{\text{susy}}_{\mathcal{A}}$ on the geometry of $\mathcal{A}$, since it also depends on the field content of the theory and on the gauge group as discussed in the previous section. 
 
If $S^{\text{susy}}_{\mathcal{A}}$ indeed represents the finite piece of entanglement entropy, we are led to conclude that supersymmetry places restrictions on the theory in such a way that the correlation of degrees of freedom across the boundary is highly insensitive to the geometry of $\mathcal{A}$. Given a fixed field content, gauge group and correlation length of the theory,
the correlations measured by $S^{\text{susy}}_{\mathcal{A}}$ are dependent only on the topology of $\mathcal{A}$ and therefore highly non-local. 

\begin{figure}
    \centering    \includegraphics[width=0.65\linewidth]{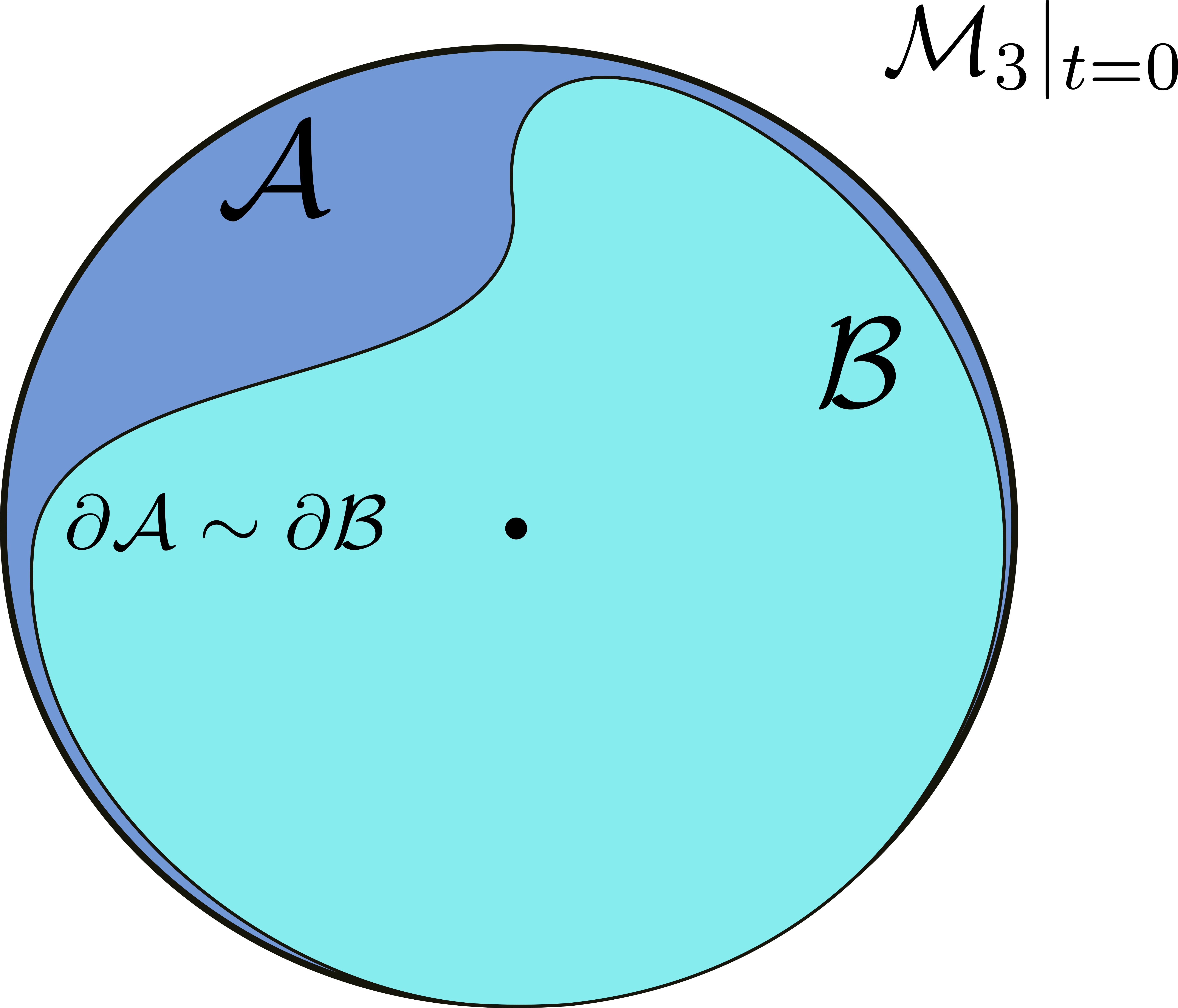}
    \caption{Two entangling regions smoothly deformed with respect to one another (denoted by $\partial\mathcal{A}\sim\partial\mathcal{B}$) and with respect to a circular region in a Cauchy slice of $\mathcal{M}_3$. Configurations of entangling regions of this form are such that quantum correlations across the boundaries $\partial\mathcal{A}$ and $\partial\mathcal{B}$ lead to $I(\mathcal{A},\mathcal{B})=0$.}
    \label{fig:adjoint_reg}
\end{figure}

Another way to differentiate the supersymmetric and the ordinary entanglement entropies is to consider mutual information, which for intersecting regions reads
\begin{equation}    I(\mathcal{A},\mathcal{B})=S_{\mathcal{A}}+S_{\mathcal{B}}-S_{\mathcal{A}\cup\mathcal{B}}-S_{\mathcal{A}\cap\mathcal{B}}\, .
\end{equation}
This quantity is cutoff-independent, and is usually taken as a measure of universal correlations captured by the entanglement entropy. If we consider $\mathcal{A}$ and $\mathcal{B}$ to be adjoint regions smoothly deformed with respect to each other and with respect to the disk $S^1$ (see Figure \ref{fig:adjoint_reg}), then all four terms above equal the disk supersymmetric entanglement entropy (in absolute value), implying that $I(\mathcal{A},\mathcal{B})=0$. The fact that the mutual information not only is invariant but actually vanishes signals suppression of quantum correlations across the boundary, seemingly originating from supersymmetry.

Another property of $S_n^{\text{susy}}$ that is topological in nature is the logarithmic divergence found in the second part of our paper. As argued in \cite{SS_Casini_2020}, the specific logarithmic divergence in (\ref{deltaI}) exists in any $d>2$ hence cannot be expressed in terms of curvature invariants on the boundary. It is also independent of specific interactions of the theory at short or large distances or mass scales as it follows from charge generators of the global symmetry group $G$. Therefore it is tempting to identify it with topological entanglement entropy \cite{Kitaev_2006}, which in the case of finite group $G$ works out as
\begin{equation}
    |G|=\sum_rd^2_r\Rightarrow \Delta I=2\log\mathcal{D}\text{,}
\end{equation}
where $\mathcal{D}=\sqrt{\sum_rd_r^2}$ is the total quantum dimension. This leads to a finite term in the difference in entanglement entropies $\Delta S=\frac{\Delta I}{2}=\log\mathcal{D}$, resulting from a negative contribution $-\log\mathcal{D}$ to $S_{\mathcal{O}}$. 
\vspace{4mm}

\noindent\textbf{Logarithmic coefficients. } The coefficient of the logarithmic divergence we found depends only on the rank of the gauge group, however, it disagrees with the value $-\frac{\text{dim}(\mathfrak{g})}{2}$ predicted in \cite{SS_Casini_2020} (the fact that the coefficients agree for $\mathfrak{g}=\mathfrak{u}(1),\mathfrak{u}(2)$ may simply be a coincidence). It was also not clear how to interpret the universal contribution to the entanglement from each $U(N)$ chiral and vector multiplet for $N\geq 2$. In fact, a precise explanation of the origin of the $-\frac{1}{2}\log 2$ contribution from a $U(1)$ or even a free chiral multiplet found already in \cite{SUSYRenyi_Nishioka_2013} is lacking. It seems that this value corresponds to a Shannon entropy of finitely many degrees of freedom resulting from the cancellation of infinitely many propagating modes in the one-loop determinants due to supersymmetry. An interpretation of this sort would clarify the values which are obtained for higher rank gauge groups. 
Our result could be reproduced by countour integration techniques involving JK residues presented also in \cite{AtwistClosset:2017zgf}. We leave this to future work. 

As an aside, it is interesting to note that the finite terms in our result arise from an integration over the gauge moduli where the integrand is written exclusively in terms of UV divergent terms produced by the heat kernel expansion. This is reminiscent of the method in \cite{Wilczek_Hertzberg_2011}, where a mass-dependent finite term is extracted from the heat kernel expansion but does not result from this expansion directly.

To relate these results to the first part of the paper, we observe that (\ref{susyRenyi}) contains physical information about the underlying theory, especially on its field content and gauge group even though its geometric dependence on $\mathcal{A}$ was found to be highly restricted.
\vspace{4mm}

\noindent\textbf{Holography.} One can test our results in holography using the holographic dual of $3d$ $\mathcal{N}=2$ $U(N)_k\times U(N)_{-k}$ ABJM theory \cite{ABJMAharony_2008} which is given by type IIA supergravity on AdS$_4\times\mathbb{CP}^3$. One could also consider computing the holographic entanglement entropy in this supergravity theory and matching it to the predictions for the UV divergent terms of the entanglement entropy of ABJM at large $N$ similarly to what was done in \cite{entropy_Fursaev_2012} for $4d$ $\mathcal{N}=4$ SYM (the universal terms were already matched in \cite{MarinoDrukker_2011}). 
\vspace{0.4cm}

\noindent\textit{Acknowledgements.}
This work is based on PM's Master's thesis project supervised by UG at the Institute for Theoretical Physics, Utrecht University. 
We are grateful to César Agón and Horacio Casini for useful discussions and especially to Itamar Yaakov for detailed explanations of their work and Abram Akal for collaboration at the early stages of the project and continuous help with supervision of PM. PM's Master's thesis can be found at https://studenttheses.uu.nl/handle/20.500.12932/47163, and contains more background material, as well as other analyses which we had to omit here. UG is supported by the Netherlands Organisation for Scientific Research (NWO) under the VICI grant VI.C.202.104.

%% file: sections/appendix1.tex
\section{THF on the three-sphere}\label{appendix1}

We first want to express the metric on $S^3_b$ in terms of the metric of a Seifert manifold,
\begin{equation}\label{THFF}   
\begin{aligned}
&ds_{\mathcal{M}_{3}}^{2}=\eta^{2}+c^{2}(z,\bar{z})dzd\bar{z}\text{,}\\
&\eta=\eta_{\mu}dx^{\mu}=d\psi+h(z,\bar{z})dz+\bar{h}(z,\bar{z})d\bar{z}\text{.}
\end{aligned}
\end{equation}
Using Hopf coordinates on $S^3$, we start from
\begin{equation}\label{Hopf}
ds^{2}_{S^3}=d\theta^2+b^2\sin^{2}\theta d\tau^{2}+b^{-2}\cos^{2}\theta d\varphi^{2}\text{,}
\end{equation}

The adapted coordinates on the THF $(\psi,z,\bar{z})$ can be expressed in terms of $(\theta,\tau,\varphi)$ through \cite{reviewClosset_2019}
\begin{equation}
z=f(\theta)e^{i\phi}\text{,}\,\,\,\,\,\begin{pmatrix}\phi\\
\psi
\end{pmatrix}=\begin{pmatrix} b^{-1} & -b\\
\frac{1}{2} & \frac{1}{2}
\end{pmatrix}\begin{pmatrix}\varphi\\
\tau
\end{pmatrix}\text{,}
\end{equation}
We now want to find the functions $f(\theta)$, $h(z,\bar{z})$ and $c(z,\bar{z})$ such that (\ref{THFF}) and (\ref{Hopf}) agree. Using the reparametrization above, it is simple but tedious to show that the functions $c$ and $f$ expressed in Hopf coordinates take the form
\begin{equation}
    \begin{cases}
    c=\cos\theta\\
    f=2\sin\theta\text{.}
\end{cases}
\end{equation}

This determines the THF and the adapted metric of the Seifert fibration
describing the manifold $S^{3}$.

We now express $f(\theta)$ and $c(\theta)$ in terms of flat space coordinates at $t=0$ related to $\theta$ through the CHM map. It is again tedious to show that we have
\begin{equation}
\begin{aligned}
    &\sin^2\theta\Big|_{t=0} =\frac{1}{\cosh^2u}\Big|_{t=0}=\left(\frac{R^2-\rho^2}{R^2+\rho^2}\right)^2\Rightarrow\\
    \Rightarrow &\cos^2\theta\Big|_{t=0}=\frac{4R^2\rho^2}{(R^2+\rho^2)^2}\text{.}
    \end{aligned}
\end{equation}

%% file: sections/appendix2.tex
\section{Vanishing contributions from potential terms} \label{appendix2}

Here we show that the potential terms in the Lagrangians on the branched sphere do not contribute to the heat coefficients, by showing that their integral over the regularized branched sphere contains no terms linear in $1-n)$ in the $n\to 1$ limit.
\vspace{0.4cm}

\noindent\textit{Scalar potential. }It is crucial to note that the
potentials we will consider here depend 1) on the regularization of the manifold through $f_{\delta}(\theta)$; or 2) explicitly on $n$, due to the presence of $A_{\mu}^{(R)}$. Therefore, terms $\propto(1-n)$ could arise when integrating $V$
over $\tilde{S}_{n}^{3}$ in (\ref{bulk_coeff}), which should
be included in the surface coefficients. 

Because our potential does not contain any derivatives of $f_{\delta}$,
terms due to 1) do not appear; see Appendix C of \cite{SUSYRenyi_Nishioka_2013} to understand how such terms arise from derivatives of $f(\delta)$. This allows us to simply take the
limit $\delta\to0$. The result of the integration over $\tilde{S}_{n}^{3}$
is therefore equal to the result on $S_{n}^{3}$, and no surface terms
occur. However, there is still a dependence on $(1-n)$ to second
order due to 2). One might think this would give contributions not
to $S_{\mathcal{A}}$, but rather to $S_{\text{\ensuremath{\mathcal{A}}}}^{(n)}$.
However, this contribution actually vanishes. It is given by (including the regularized configuration for $A_{\mu}^{(R)}$)
\begin{equation}\label{a_1_vanishes}
\begin{aligned}
&\int_{\tilde{S}_{n}^{3}}\sqrt{g}\frac{1}{n^{2}\sin^{2}\theta}\left(n\sqrt{f_{\delta}}-1\right)^{2} \\ \simeq &  \int_{0}^{\epsilon}d\theta\frac{1}{n^{2}\theta\sqrt{f_{\delta}}}\left(n\sqrt{f_{\delta}}-1\right)^{2}+\int_{\epsilon}^{\frac{\pi}{2}}d\theta\cot\theta(n^{2}-1)=0\text{,}
\end{aligned}
\end{equation}
where we used that $\sqrt{f_{\delta}(\theta=\delta)}=1$
and for $\theta\to0$ the integrand behaves as $\frac{1}{\theta}\left(n\sqrt{f_{\delta}}-1\right)^{2}=\mathcal{O}(\theta)\to 0$.

Since contributions to $a_1^{\Sigma}$ would come from extracting terms $\propto(1-n)$ in (\ref{a_1_vanishes}), we conclude that there are no surface contributions from the potential in the scalar Lagrangian
and only the Ricci-like surface terms (\ref{surf_scalar}) remain.

\vspace{0.4cm}
\noindent\textit{Spinor potential. }A comment on the spinor operator is in order here. In \cite{cones_Fursaev_1997}
the heat coefficients for the free fermion Laplacian are determined by taking its
square, 
\begin{equation}
    (-i\slashed{\nabla})^{2}=-\nabla^{\mu}\nabla_{\mu}+\frac{1}{4}R\text{,}
\end{equation}
and dividing the heat coefficients of the resulting operator by $2$
(since the logarithm of the eigenvalues of $-i\slashed{\nabla}$ is
half of the logarithm of eigenvalues of $(-i\slashed{\nabla})^{2}$).
Any other terms that we add to the spinor operator can be treated
as explained above, by including their contribution to the coefficients
in the term $-\int_{\mathcal{M}}\text{Tr}_{i}(X^{(j)})$.

The trace over spinor indices that arises when computing
$\text{Tr}_{i}X^{(j)}$ is

\begin{equation}
(\gamma^{\tau})_{\alpha}^{\,\,\,\alpha}=\varepsilon^{\beta\alpha}(\gamma^{\tau})_{\alpha\beta}=2in\cot\theta\sin(\tau+\phi)\text{,}
\end{equation}
and this gives
\begin{equation}
\begin{aligned}\int_{\tilde{S}_{n}^{3}}\text{Tr}_{i}(X^{(j)}) & =-i\int_{0}^{\frac{\pi}{2}}d\theta\int_{0}^{2\pi}d\phi\int_{0}^{2\pi}d\tau\frac{\sin\theta\cos\theta}{n\sqrt{f_{\delta}}}\times\\
& \,\,\,\,\,\,\,\,\times\cot\theta\sin(\tau+\phi)\frac{n\sqrt{f_{\delta}}-1}{2}=0\text{.}\end{aligned}
\end{equation}
Again, no surface terms arise due to the potential. 

%% file: sections/appendix3.tex
\section{Spinor heat coefficients at finite coupling}\label{appendix3}

The quadratic operator of the fermion field in the chiral multiplet on $S^3$ is now $\mathcal{D}{}_{\psi}\equiv-i\slashed{\nabla}-i\sigma_{0}$, where the partition function is obtained through an overall integration over $\sigma_0$. In the free case we have related the coefficients of the Dirac operator with no mass terms, $-i\slashed{\nabla}$, to those of
$(-i\slashed{\nabla})^{2}=-\nabla^{2}+\frac{1}{4}R$. In the presence of a mass term this procedure is not so simple, since squaring $-i\slashed{\nabla}-i\sigma_{0}$
would give us both second and first order terms in $\slashed{\nabla}$.
We could try to solve for the coefficients of
\begin{equation}
\begin{aligned}(-i\slashed{\nabla})^{2}+\sigma_{0}^{2} & =(-i\slashed{\nabla}-i\sigma_{0})(-i\slashed{\nabla}+i\sigma_{0})\text{,}\end{aligned}
\end{equation}
but then we would not be able to relate them to the heat kernel coefficients
of $\mathcal{D}_{\psi}$ alone. Instead, we make the change of integration variables
$\sigma_{0}\to i\sigma_{0}$, and the operator becomes $\tilde{\mathcal{D}}_{\psi}\equiv-i\slashed\nabla-\sigma_{0}\textbf{1}$.
This is no longer an operator with purely imaginary eigenvalues; we
can write
\begin{equation}
\begin{aligned}
\tilde{\mathcal{D}}_{\psi}^{\dagger}\tilde{\mathcal{D}}_{\psi}&=(-i\slashed{\nabla}-\sigma_{0})(i\slashed{\nabla}-\sigma_{0})=-(i\slashed{\nabla})^{2}+\sigma_{0}^{2}\\&=\nabla^{2}-\frac{1}{4}R+\sigma_{0}^{2}\text{,}
\end{aligned}
\end{equation}
The eigenvalues
of this operator are the square of the absolute value of the eigenvalues
of $\tilde{\mathcal{D}}_{\psi}$ . We then take the heat kernel coefficients
of $\tilde{\mathcal{D}}_{\psi}$ as half of the heat kernel coefficients
of $\tilde{\mathcal{D}}_{\psi}^{\dagger}\tilde{\mathcal{D}}_{\psi}$.
Explicitly, the identification we are making in order to compute the
heat kernel coefficients is
\begin{equation}
\lambda_{\tilde{\mathcal{D}}_{\psi}}^{*}\lambda_{\tilde{\mathcal{D}}_{\psi}}\longleftrightarrow(\lambda_{\tilde{D}_{\psi}})^{2}\text{.}
\end{equation}
These two quantitites differ
by a complex phase, $\lambda^{*}\lambda=e^{-2i\text{arg}(\lambda)}\lambda^{2}$,
meaning that the difference between applying $\log\det$ to $\tilde{\mathcal{D}}_{\psi}^{\dagger}\tilde{\mathcal{D}}_{\psi}$
and to $\tilde{\mathcal{D}}_{\psi}$ is, besides the factor of $2$,
a purely imaginary additive quantity which we are not concerned with and may safely disregard. The spinor heat kernel coefficients then become
\begin{equation}
\begin{aligned}a_{1,\psi}^{\text{bulk}}(S_{n}^{3}) & =-\frac{2^{[3/2]}}{2}n\int_{S^{3}}\sqrt{g}d^{d}x\left(-\frac{1}{6}R-\left(-\frac{1}{4}R+\sigma_{0}^{2}\right)\right)\\
 & =n\text{Vol}(S^{3})\left(\frac{1}{2l^{2}}+\sigma_{0}^{2}\right)\text{.}
\end{aligned}
\end{equation}

We now consider the gaugino sector in the non-abelian case. For the components $\lambda_{\alpha}$ of the gaugino along each root space labelled by $\alpha$, this operator is
\begin{equation}
\mathcal{D}_{\lambda_{\alpha}}=-i\slashed{\nabla}+\textbf{1}\left(-i\alpha(\sigma_{0})+\frac{1}{2}\right)\text{.}
\end{equation}
As above we now consider heat kernel coefficients of the following operator:
\begin{equation}
\begin{aligned}
\tilde{\mathcal{D}}_{\lambda_{\alpha}}^{\dagger}\tilde{\mathcal{D}}_{\lambda_{\alpha}}\equiv &\left(-i\slashed{\nabla}+\alpha(\sigma_{0})+\frac{1}{2}\right)\left(i\slashed{\nabla}+\alpha(\sigma_{0})+\frac{1}{2}\right)=\\=&\nabla^2-\frac{1}{4}R+\left(\alpha(\sigma_{0})+\frac{1}{2}\right)^{2}\text{,}
\end{aligned}
\end{equation}
obtaining
\begin{equation}
\begin{aligned}
a_{1,\lambda}^{\text{bulk}}(S_{n}^{3}) = n\text{Vol}(S^{3})\left[\frac{1}{2l^{2}}+\left(2(\sigma_{1}-\sigma_{2})^{2}+\frac{1}{2}\right)\right]\text{.}
\end{aligned}
\end{equation}

%% file: sections/appendix4.tex
\section{Some computational details on the heat coefficients}\label{appendix4}

\noindent\textit{U(1) chiral multiplet. }
We have
\begin{equation}
\begin{aligned}
&a_{1,\phi}^{\text{bulk}}(S_{n}^{3}) =na_{1,\phi}^{\text{bulk}}(S^{3})\\
 = & n\int_{S^{3}}\sqrt{g}d^{d}x\left(\frac{1}{6}R-\left(\frac{1}{2}H+\sigma_{0}\right)^{2}+H^{2}+\frac{1}{8}(R-6H^{2})\right)\\
 =& n\text{Vol}(S^{3})\left(-\frac{7}{4l^{2}}+\frac{i}{l}\sigma_{0}-\sigma_{0}^{2}\right)\text{,}
\end{aligned}
\end{equation}
where $R(S^{3})=-6/l^{2}$ has been used.
\vspace{4mm}

\noindent\textit{Two chiral multiplets in $\textbf{1}$ and $\overline{\textbf{1}}$
of $U(1)$.} Before the complexification of the $\sigma_0$ variables, the relevant coefficients are
\begin{equation}
\begin{aligned}
&a_{1,\phi_{\pm}}^{\text{bulk}}(S_{n}^{3})=n\text{Vol}(S^{3})\left(-\frac{7}{4l^{2}}\pm\frac{i}{l}\sigma_{0}-\sigma_{0}^{2}\right)\text{,}\\ & a_{1,\psi_{\pm}}^{\text{bulk}}(S_{n}^{3})=a_{1,\psi}^{\text{bulk}}(S_{n}^{3})\text{,}
\end{aligned}
\end{equation}
where $\phi_\pm$ denotes the field in the $\textbf{1}$ or $\overline{\textbf{1}}$ representation.
\vspace{4mm}

\noindent\textit{$U(2)$ chiral multiplet.} Starting with the gauge group $U(2)$, we expand the one-loop determinants  in heat kernel coefficients.
The determinant from the chiral sector factorises into a product of determinants depending on each of the weights of $\mathfrak{u}(2)$, and the expression for $\exp\left(\log Z_{\text{matter}}^{\text{1-loop}}(\sigma_{0},n)\right)$ becomes
\begin{equation}
\prod_{i=1}^{2}\exp\left(na\left(-\frac{11}{4l^{2}}+2(\sigma_{i}-\frac{1}{4l})^{2}\right)\frac{l}{\epsilon}+(1-n)\frac{3\sqrt{\pi}}{8}\frac{l}{\epsilon}\right)\text{.}
\end{equation}

There is now a potential in $\mathcal{L}_{\text{gauge}}$ which must be accounted for in the corresponding heat kernel coefficients. This is due to the non-trivial root spaces of $U(2)$ along which the fields in the gauge multiplet are expanded (see \cite{Kapustin_Yaakov_exact}). In this case we have $N_{\alpha}=2$ and $N_{\text{Cartan}}=2$, so that
\begin{equation}
\begin{aligned}
&a_{1,B}^{\text{bulk}}(S_{n}^{3})  =na_{1,B}^{\text{bulk}}(S^{3})\\
 =& nN^{(1)}\int_{S^3}\sqrt{g}\left((N_{\text{Cartan}}+N_{\alpha})\frac{1}{6}R-\sum_{\alpha}\alpha(\sigma_{0})^{2}\right)\\
 =& -n\text{Vol}(S^{3})\left(6+4(\sigma_{1}-\sigma_{2})^{2}\right)\text{.}
\end{aligned}
\end{equation}

The coefficients for the gaugino sector can be obtained through a similar analysis as in the abelian case, and we leave it to Appendix \ref{appendix3}. The terms from the bulk coefficients contributing to the $\sigma$
integral amount to
\begin{equation}
\begin{aligned} 
& \int_{-\infty}^{+\infty}d\sigma_{1}d\sigma_{2}(\sigma_{1}-\sigma_{2})^{2}\times\\ &\,\,\,\,\,\,\,\,\times\exp\left[na{(4\pi)^{3/2}}\left(-4(\sigma_{1}-\sigma_{2})^{2}-2\sigma_{1}^{2}-2\sigma_{2}^{2}\right)\frac{l}{\epsilon}\right]\text{.}
\end{aligned}
\end{equation}
Making the change of variables
$\sigma_{+}\equiv\sigma_{1}+\sigma_{2}$, $\sigma_{-}\equiv\sigma_{1}-\sigma_{2}$,
we find
\begin{equation}\label{integral}
\begin{aligned} & \int_{-\infty}^{+\infty}d\sigma_{+}d\sigma_{-}\sigma_{-}^{2}\exp\left[na\left(-5\sigma_{-}^{2}-\sigma_{+}^{2}\right)\frac{l}{\epsilon}\right]\\
= 
& \frac{\pi}{2\cdot 5^{3/2}}\left(\frac{\epsilon}{nal}\right)^2 = \frac{8}{5^{3/2}n^2l^4}\left(\frac{\epsilon}{l}\right)^2\text{.}
\end{aligned}
\end{equation}
Upon taking logarithms to obtain the entanglement entropy, we arrive at (\ref{U(2)_chiral}).
\vspace{4mm}

\noindent\textit{Two chiral multiplets in the \textbf{2} and $\overline{\textbf{2}}$ representations of $U(2)$.} Gathering the bulk coefficients for each of chiral multiplets in conjugate representations of $U(2)$ as in the abelian example, the integral we must perform is
\begin{equation}\label{U(2)integral}
 \int_{-\infty}^{+\infty}d\sigma_{+}d\sigma_{-}\sigma_{-}^{2}\exp\left[na\left(-6\sigma_{-}^{2}-2\sigma_{+}^{2}\right)\frac{l}{\epsilon}\right]\text{,}
\end{equation}
which leads to (\ref{2U(2)_chirals}).
\vspace{4mm}

\noindent\textit{$U(N)$ gauge group with $N_f$ flavours. }Here we simply need to notice that we can introduce the variables
\begin{equation}   \sigma_{ij}^{\pm}\equiv\sigma_i\pm\sigma_j\text{,}\,\,\,\,\,\,\,\,\,1\leq i\neq j\leq N\text{,}
\end{equation}
and that each term of the form $\sigma_i^2$ can be written as
\begin{equation}
    \frac{N-1}{N-1}\sigma_i^2=\frac{1}{N-1}\sum_{j\neq i}\left(\frac{\sigma_{ij}^++\sigma_{ij}^-}{2}\right)^2\text{.}
\end{equation}
Then, the resulting integration over $\mathfrak{u}(N)$ is equivalent to having an integral of the type (\ref{integral}) for each pair $i,j$. In particular, the number of such integrals we get is
\begin{equation}
\frac{N_{\alpha}}{2}=\sum_{k=1}^{N-1}(N-k)=\frac{N(N-1)}{2}\text{,}
\end{equation}
taking the form
\begin{widetext}
\begin{equation}\label{U(N)integral}
\begin{aligned}
    &\int_{\mathfrak{u}(N)}[d\sigma]\exp\left\{na\left(-4(\sigma_{ij}^-)^2-\frac{2}{N-1}\left(\frac{\sigma_{ij}^++\sigma_{ij}^-}{2}\right)^2-\frac{2}{N-1}\left(\frac{\sigma_{ij}^+-\sigma_{ij}^-}{2}\right)^2\right)\frac{l}{\epsilon}\right\}=\\
    =\,\,&\frac{1}{\text{Vol}(U(N))}\left\{\frac{8}{n^2l^4}(4N-3)^{-\frac{3}{2}}(N-1)^2\left(\frac{\epsilon}{l}\right)^2\right\}^{\frac{N(N-1)}{2}}\text{,}
\end{aligned}
\end{equation}
\end{widetext}
where the measure on the Lie algebra is
\begin{equation}
    \int_{\mathfrak{u}(N)}[d\sigma]\equiv\frac{1}{\text{Vol}(U(N))}\int_{-\infty}^{+\infty}\prod_{1\leq i< j\leq N}d\sigma_{ij}^+d\sigma_{ij}^-(\sigma_{ij}^-)^2\text{.}
\end{equation}